\documentclass[lettersize,journal]{IEEEtran}
\usepackage{amsmath,amsfonts}
\usepackage{algorithmic}
\usepackage{algorithm}
\usepackage{array}
\usepackage[caption=false,font=normalsize,labelfont=sf,textfont=sf]{subfig}
\usepackage{textcomp}
\usepackage{stfloats}
\usepackage{url}
\usepackage{verbatim}
\usepackage{graphicx}
\usepackage{cite}
\usepackage{threeparttable}
\usepackage[scaled=0.8]{beramono}
\usepackage[T1]{fontenc}
\usepackage{shortcuts}
\usepackage{amsfonts}         % Redefined in newtxmath.sty
\usepackage{amssymb} 	% For the purpose to use symbol like leqslant.
\usepackage{amstext}

\usepackage{amsthm}		% For the purpose of writing definition and/or proof of theorem
\usepackage{framed}
\usepackage{xcolor}
\usepackage{url}

\usepackage{breakurl}
\usepackage{makecell}
\usepackage{multirow}
\usepackage{tablefootnote}

\usepackage{array}
\newcolumntype{H}{>{\setbox0=\hbox\bgroup}c<{\egroup}@{}}
\usepackage[linesnumbered,ruled,algo2e]{algorithm2e}
\usepackage{framed}
\usepackage{listings}
\usepackage{comment}
\usepackage{threeparttable}
\usepackage{wrapfig}

\usepackage{caption}

\usepackage{lipsum} 
\usepackage{scalerel}
\usepackage{float}

\usepackage{enumitem}	% For the purpose of managing indent and margin of itemize environment.
\setlist{leftmargin=1.25em}
\usepackage{setspace}

\usepackage{footmisc}
\begin{document}

\title{Assessing Privacy Compliance of Android Third-Party SDKs} 
%\title{A Large-Scale Privacy Assessment of Android Third-Party SDKs} 
%\title{A Large-Scale Privacy Measurement of Android Third-Party SDKs} 
%\title{A Large-Scale Measurement of Privacy Compliance in Android Third-Party SDKs} % Proposed by baigd on 30 Jan
% \title{Assessing the Compliance of Privacy Disclosure in Android Third-Party SDKs} % Proposed by cyan on 18 Oct

\author{Mark~Huasong~Meng,
	Chuan~Yan,
	Qing~Zhang,
	Zeyu~Wang,
	Kailong~Wang,\\
	Sin G.~Teo,
	Guangdong~Bai,
	and~Jin~Song~Dong
	\thanks{M. H. Meng is with Technical University of Munich, Germany. 
		C. Yan is with the University of Queensland, Australia.
		G. Bai is with the University of Queensland, Australia, and National University of Singapore, Singapore.
		Q. Zhang and Z. Wang are with ByteDance Group, China,
		K. Wang is with Huazhong University of Science and Technology, China,
		S. G. Teo is with Agency for Science, Technology and Research, Singapore,
		and J. S. Dong is with National University of Singapore, Singapore.}}

\maketitle

\begin{abstract}
	Third-party Software Development Kits (SDKs) are widely adopted in Android app development, to accelerate development pipelines and enhance app functionality effortlessly. 
However, this convenience raises substantial concerns about unauthorized access to users' privacy-sensitive information, which could be further abused for illegitimate purposes like user tracking or monetization. 
Our study offers a targeted analysis of user privacy protection among Android third-party SDKs, filling a critical gap in the Android software supply chain. 
It focuses on two aspects of their privacy practices, including \emph{data exfiltration} and \emph{behavior-policy compliance}~(or \emph{privacy compliance}), utilizing techniques of taint analysis and large language models. 
It covers \allsdk widely-used SDKs from two key SDK release platforms, the official one and a large alternative one. 
From them, we identified 338 instances of privacy data exfiltration. 
On privacy compliance, our study reveals that more than 30\% of the examined SDKs fail to provide a privacy policy to disclose their data handling practices. 
Among those that provide privacy policies, 37\% of them over-collect user data, and 88\% falsely claim access to sensitive data. 
We revisit the latest versions of the SDKs after 12 months. Our analysis demonstrates a persistent lack of improvement in these concerning trends. Based on our findings, we propose three actionable recommendations to mitigate the privacy leakage risks and enhance privacy protection for Android users.
Our research not only serves as an urgent call for industry attention but also provides crucial insights for future regulatory interventions.

\end{abstract}

\begin{IEEEkeywords}
	mobile, privacy assessment, taint analysis.
\end{IEEEkeywords}

\section{Introduction}

The advent and ubiquity of third-party Software Development Kits (SDKs) in Android app development herald a profound shift in the mobile software development paradigm, transforming efficiency and accelerating innovation. These SDKs provide pre-built reusable code that offers a broad spectrum of functionalities, from fundamental features like user authentication~\cite{authe_lib} and data encryption~\cite{encryption_lib} to specialized components like analytics~\cite{analytics_lib}, advertisement delivery~\cite{advertisement_lib}, and user interface enhancements~\cite{UI_lib}. The open-source nature and global reach of the Android ecosystem further fuel the proliferation of these SDKs. 

Integrating an SDK into an app is often as simple as declaring dependencies in the build configuration file, thus drastically reducing development time, and allowing developers to focus on app-specific features. 
%Nowadays, the majority of Android apps incorporate a multitude of third-party SDKs, with an average app integrating approximately 8.6 SDKs~\cite{salza2018,zhang2018detecting}.
Nowadays, the majority of Android apps incorporate multiple third-party SDKs, with an average app integrating approximately 8.6 SDKs~\cite{salza2018,zhang2018detecting}.
However, this widespread dependency also introduces significant security and privacy concerns. The complexity of the software supply chain and the opaque data handling practices of some SDKs can lead to privacy risks and security vulnerabilities, either due to malicious intent or unintentional oversights. These can be exploited by malevolent actors or lead to inadvertent privacy violations, as evidenced by recent high-profile incidents related to third-party SDKs~\cite{priv_leak_blackkite,priv_leak_mahajan}.

The privacy issues caused by SDKs could be further magnified in the pre-installed apps on Android devices. 
These apps are bundled by device manufacturers or network carriers, and therefore, often have an extremely great user population. 
They come with high privileges and are typically non-removable, such that an SDK embedded by them can stealthily perform higher-privileged operations that normal apps are unable to.
An illustrative example is an app named ``\emph{Mobile Services Manager}'' pre-installed on certain Android devices, which was found to download apps without user consent due to a malicious third-party SDK embedded~\cite{htfma2023msm,mollah2022mobile}.

Assessing privacy practices of SDKs includes two key aspects, i.e., \emph{data exfiltration} and \emph{behavior-policy compliance}~(or \emph{privacy compliance}). 
The data exfiltration involves the behaviors of locating sensitive data in publicly accessible places, such as the publicly accessible directories in file systems and the Internet. 
%Assessing \emph{privacy data exfiltration} in third-party SDKs presents a complex landscape fraught with challenges.
The opacity of SDKs' operations and their ``black-box'' nature hinders meaningful assessments, leaving both end-users and app developers in the dark and elevating risks.
Dishonest or malicious SDKs may disguise as their embedding apps to extensively collect privacy-sensitive data, while the end-users are unable to distinguish which party requests their privacy data.
Incidents like the X-Mode controversy~\cite{keegan2022sold}, where user locations were tracked by over 100 apps through embedded SDKs, underscore this point and serve as a stark reminder of the challenges associated with preserving user's privacy in third-party SDKs.

On the regulatory front, data protection regulations have been put in place globally, such as the General Data Protection Regulation (GDPR)~\cite{gdpr2016} and the California Consumer Privacy Act (CCPA)~\cite{ccpa2023california}, which impose a stringent requirement on data processors. 
In response to this, software developers are required to provide a privacy policy to explicitly disclose what personal information is collected and how these data are handled by their software. 
Dissecting whether the data handling behaviors of SDKs comply with the claims in their privacy policies, which we define as behavior-policy compliance~(or privacy compliance), becomes an urgent task. 

\paragraph{Our Work} In this work, we conduct a large-scale study on the privacy practices of existing third-party SDKs. 
Our study covers \allsdk prevalently-installed SDKs from two mainstream release platforms, including Google Play SDK Index~\cite{google_play_sdk_index} and CAICT SDK Index~\cite{caict}. 
The former is the official source of Android SDKs as a part of Google Play infrastructure, which lists the most widely used commercial SDKs in the Android app ecosystem~\cite{rodriguez2024sharing}. 
The latter is claimed to be the largest release platform of mobile SDKs in China, serving as a complementary source of SDKs due to the absence of Google's service in the Chinese market.
Based on the statistical data provided by the two platforms, over 100 collected SDKs have been embedded by at least 100 apps with over 10 million installations, demonstrating the broad impact of this study.

Our study comprises two main strategies, i.e., conducting a data exfiltration assessment through static program analysis, and assessing privacy compliance through interpreting privacy policies. 
Specifically, we conduct a static taint analysis for tracking the flow of privacy-sensitive data within the SDK.
We explore the traces that read and share privacy-sensitive data in public spaces, and accordingly investigate whether these traces can lead to potential privacy-sensitive data exfiltration. 
For the privacy compliance assessment, we employ a large language model (LLM) to analyze the SDKs' privacy policies and thereby identify what data is requested, cross-check the data claimed in privacy policies with the data access behaviors caught during the taint analysis, and determine if an excessive collection or over-claiming issue exists.

Our study reveals critical privacy shortcomings in Android third-party SDKs. Astonishingly, \taintallsdk traces are found to read privacy data and share them in publicly accessible places.
Among these traces, \correcttaintallsdk are confirmed to be subject to data leakage. 
Only 109 out of the \allsdk examined SDKs provide privacy policies, among which approximately 37\% over-collect private data and over 88\% falsely claim their data collection scopes, seriously violating global data protection norms. 
Our re-inspection after 12 months shows no meaningful improvement in these practices over time. 
We also uncover a novel system settings injection issue: some malicious SDKs exploit pre-installed apps to share privacy-sensitive data into Android OS's public system settings, effectively creating a unique user-unresettable identifier (UUI) accessible to any app. This unprecedented finding highlights an urgent need for improvement in SDK development practices, permission management, and privacy violation detection. To this end, we further propose three key strategic recommendations as mitigation. 

\paragraph{Contributions} Our work makes the following key contributions:
\begin{itemize}

\item \textbf{Detecting Potential Privacy Data Exfiltration on a Large Scale.} We collect \allsdk SDKs from two mainstream repositories and leverage taint analysis to track sensitive data flow, thereby exploring their potential data exfiltration. 
We find \correcttaintallsdk data exfiltration evidenced by sharing obtained privacy data into public spaces. Our work highlights the substantial privacy risks stemming from the widespread use of third-party SDKs in the Android ecosystem, unveiling complex and often opaque data-handling practices.

\item \textbf{Revelation of Privacy Compliance Landscape at the SDK level.} Our study uncovers a startling level of data over-collection among third-party SDKs, with an estimated 37\% of SDKs engaging in this practice. Through our re-inspection, we depict the trends of SDKs' practices in collecting and sharing privacy data, and underscore the urgency of tackling privacy risks in the Android ecosystem at a fine-grained SDK level. 

%\item \textbf{Discovery of Amplified Privacy Risk caused by Pre-installed Privileges.} % Remove the word 'amplified' for typesetting purpose
\item \textbf{Discovery of Privacy Risk caused by Pre-installed Privileges.} 
We find a system settings injection issue in the pre-installed apps, in which a malicious SDK can excessively read and share privacy data into the system settings. 
This enables an arbitrary pre-installed app with the malicious SDK to create uniquely unresettable identifiers and share them, violating the relevant regulations about privacy protection. 
%This can \highlight{turn any data into a de facto UUI} \baigd{any data? or arbitrary app with the SDK can set UUI? }and therefore, violates the relevant regulations about privacy protection. 
It is the first time such issues have been revealed and systematically investigated.
\end{itemize}

%\paragraph{Open Science Commitment}
%We will open source all artifacts involved in this study upon paper acceptance, including a dataset of Android SDKs, our code for privacy policy inference, the configuration scripts for the static analysis, and the detailed statistics of our assessment results. 
%\input{sections/background}
% !TeX root = ../main.tex
\section{Backgrounds and Problem Definitions}

\subsection{GDPR and Privacy Compliance}
The General Data Protection Regulation (GDPR)~\cite{gdpr2016}, a landmark legislation enacted by the European Union (EU) in 2018, has profoundly transformed the landscape of data privacy laws within the region. 
Under GDPR, entities processing personal data, defined broadly as any data enabling the direct or indirect identification of individuals, are subjected to stringent regulations that underscore the importance of ``\emph{lawfulness, fairness, and transparency}''. Furthermore, GDPR provides explicit requirements for obtaining valid consent, necessitating that it is freely given, specific, informed, and unambiguous.

Privacy compliance, in the context of GDPR, necessitates a thorough examination of privacy policies to ensure adherence to the GDPR's comprehensive principles and stipulations. It entails a meticulous inspection of data processing activities, thereby ensuring data is acquired and utilized only for explicit, legitimate purposes. Furthermore, the principle of ``\emph{data minimization}'' obliges organizations to limit their data processing activities to the minimum necessary, thereby reinforcing the importance of data accuracy and relevance. This process also extends to an examination of storage practices, necessitating the validation that data is retained only for as long as necessary and is safeguarded against unlawful processing, inadvertent loss, or damage. Privacy compliance, therefore, functions as a crucial mechanism to enforce accountability, reinforcing the necessity for organizations to demonstrate, unequivocally, their adherence to GDPR principles.

\subsection{Android SDKs and Data Collection Practices}
Android third-party SDKs represent a crucial component within the broader ecosystem of Android app development. These software SDKs offer pre-written code to app developers, enabling the efficient integration of different features and functions within their apps, ranging from user interface elements to complex data analytics capabilities. 
% By leveraging these SDKs, developers can expedite the application development process, avoiding the necessity to ``reinvent the wheel'' for common functionalities. Consequently, these SDKs have become an integral part of the Android development framework, enhancing productivity and fostering innovation through code reuse.
Data collection practices associated with Android third-party SDKs have been a subject of scrutiny within academic and industry communities. It has been observed that some SDKs, while providing utility functions, also engage in extensive data collection activities without explicit acknowledgment or consent received from the end-users. 
% These practices tend to infringe upon user privacy at times, collecting sensitive information such as device identifiers, location data, and personal preferences. 
Furthermore, there are instances where data is transmitted to remote servers, raising concerns about data security and potential misuse. 
% In light of this, there is an increasing emphasis on implementing robust privacy and security measures, such as privacy-preserving APIs and data anonymization techniques, within these SDKs to ensure ethical and lawful data collection practices.
In light of this, there is an increasing need of a large-scale privacy assessment of SDKs. 

\subsection{Objectives and Problem Definitions}
\label{sec:problem_definitions}

This work aims to
(1) detect potential exfiltration of users' privacy data existing in the third-party SDKs, 
(2) explore the privacy compliance of developers in the SDK release, and thereby, 
(3) unveil the landscape of privacy protection in the Android ecosystem at the SDK level.
To this end, we propose a compliance model to systematically assess the privacy practices of Android SDKs regarding personal information collection. 
We let $\mathbb{D}$ represent a set of diverse data considered as user privacy, and let $\mathbb{S}$ be the set of publicly released third-party SDKs. Then we define what privacy data an SDK claims to collect and actually collects as follows. 

%\paragraph{Data Collection Disclosure ($\mathcal{C}$)}
%We let $\mathcal{C}$ denote the disclosure of an SDK $\mathbf{s} \in \mathbb{S}$. Specifically, $\mathcal{C}_{s}$ is the set of privacy data \underline{c}laimed in the privacy policy of the SDK $\mathbf{s}$.

\paragraph{Compliance Disclosure ($\mathcal{C}$)}
We let $\mathcal{C}$ denote the \underline{c}ompliance disclosure of an SDK $s \in \mathbb{S}$. Specifically, it defines the set of privacy data claimed to access in the privacy policy of the SDK $s$. It shall also be assumed not to leak any privacy data to the public. 

\paragraph{Implementation-level Practice ($\mathcal{P}$)}
The compliance \underline{p}ractice of an SDK concerns two types of operations in its implementation, i.e., collecting privacy data, and sharing the collected privacy data.
We let $\mathcal{R}_{s}$ denote the scope of privacy data that would be \underline{r}ead/collected by an SDK ${s}$ based on its implementation, in which $\mathcal{R}_{s} \subseteq \mathbb{D}$.
We also let $\mathcal{U}_{s}$ denote the collected privacy data involved in any \underline{u}ploading/sharing operations, given that $\mathcal{U}_{s} \subseteq \mathcal{R}_{s}$.
Thus, we have the compliance practice of an SDK ${s}$ as a tuple of its collecting and sharing behaviors, written as $\mathcal{P}_{s} = (\mathcal{R}_{s}, \mathcal{U}_{s})$.

Next, we formalize three types of compliance issues concerned in our investigation.

\paragraph{Type I: Privacy Leakage} %($\nvdash_{\mathcal{P}} \mathcal{C}_{s}$)
Sharing users' privacy in public spaces poses compliance risks, especially when the collection or the sharing operations are not performed with users' consent. Moreover, some regulations like the GDPR have explicitly prohibited any collection of UUIs, amplifying the risks of privacy sharing on the SDK side. 
Here we define that any pre-recognized privacy data ${d}$ being observed in sharing operations of an SDK ${s}$ would constitute privacy sharing risk, which we formalize as the compliance not being true (i.e., strikethrough in a unary turnstile, $\nvdash$) due to implementation-level practice $\mathcal{P}$. We detail this as follows:

\vspace{-0.2cm}
\begin{equation}
	\begin{aligned}
		\label{equ:risk1}
		\exists s \in \mathbb{S}, \exists d \in \mathbb{D}, d \in \mathcal{U}_{s} \Rightarrow \,\,\nvdash_{\mathcal{P}} \mathcal{C}_{s}.
		% \exists s \in \mathbb{S}, \exists d \in \mathbb{D}, d \in \mathcal{U}_{s} \Rightarrow \mathbf{C}_{s} \nvdash_{\mathcal{U}} \mathcal{P}.\mathcal{U}_{s}.
	\end{aligned}
\end{equation}

Besides improper sharing, recent developments in privacy protection regulations also pose stricter requirements in even reading privacy data on personal devices. 
Developers are required to not only declare a list of privacy data to be collected but also to ensure the actual collection operations are always consistent with what has been declared. 
Next, we formalize two types of inconsistency issues that could happen in privacy data collection.

\paragraph{Type II: Excessive Collection} %($\nvdash_{\mathcal{P}}$) 
Excessive collection concerns the privacy practice of an SDK that attempts to read more types of privacy data than what it requests/claims in its privacy policy. 
Excessive collection infringes on users' right to be informed as the SDK collects privacy data without users’ awareness and consent.
We define this type of risk as follows.

\vspace{-0.3cm}
\begin{equation}
	\begin{aligned}
		\label{equ:risk2}
		\exists s \in \mathbb{S}, \exists d \in \mathbb{D}, (d \in \mathcal{R}_{s}) \land (d \notin \mathcal{C}_{s}) \Rightarrow \mathcal{R}_{s} \nvdash_{\mathcal{P}} \mathcal{C}_{s}.
	\end{aligned}
\end{equation}

We remark that Google has not set a formal definition of ``data collection'' at the SDK level as what it does for third-party apps~\cite{google2024understanding}. Thus, we refer to the guideline of GDPR Article 13~\cite{gdpr2016,wolford2024guide} and take a conservative strategy by considering data accessed by SDKs as \emph{collected} when evaluating the excessive collection, given that the data accessed by an SDK can flow into the host apps, which can then consume or send the data out. 

\paragraph{Type III: Over-claiming} %($\nvDash_{\mathcal{P}}$)
We assert that an SDK has an over-claiming issue if it is found to claim access to more types of privacy data than it actually reads. 
Over-claiming issues, although they may not substantially lead to privacy data exfiltration, seriously violate the \emph{data minimization principle} stipulated in Article 5(1)(c) of GDPR. It is defined as follows.

\vspace{-0.3cm}
\begin{equation}
	\begin{aligned}
		\label{equ:risk3}
		\exists s \in \mathbb{S}, \exists d \in \mathbb{D}, (d \notin \mathcal{R}_{s}) \land (d \in \mathcal{C}_{s}) \Rightarrow \mathcal{R}_{s} \nvDash_{\mathcal{P}} \mathcal{C}_{s}.
	\end{aligned}
\end{equation}

\subsection{Threat Model}

We consider that the exfiltration of users' privacy data caused by malicious SDKs should follow a unified routine. 
To achieve such exfiltration, the malicious SDK (the attacker) gains capabilities as follows:

\begin{itemize}
	\item The malicious SDK is released in a public platform/market and thus, can be embedded in an arbitrary app and installed on an Android device. Without causing an advantage to the attacker, that device is not rooted and is properly maintained with the latest security update installed.
	\item The malicious SDK accesses users' privacy data through the interfaces of the Android OS with permissions granted to the embedding app, i.e., all its collection of users' privacy data must be realized on behalf of the embedded app.
	\item The malicious SDK saves or shares the collected privacy data in a public space, i.e., there exists no party (including the embedding app)  colluding with the malicious SDK in sharing the collected privacy data, or saving it in private storage. 
\end{itemize}

% Below are components of the methodology section
%\input{sections/privacy-data}

% !TeX root = ../main.tex

\section{Methodology}
\label{sec:methodology}

%This section details our collection of existing Android SDKs, followed by our approach to analyzing data requests from SDKs' privacy policies, and the design and configuration of our static taint analysis to explore privacy collection and potential exfiltration from SDK libraries.

\subsection{Approach Overview}
\label{sec:overview}

\begin{figure*}[t]
	\centering
	% trim order: <left> <lower> <right> <upper>}
\includegraphics[trim=0.4cm 7.1cm 0.3cm 0cm,clip=true,width=0.85\linewidth]{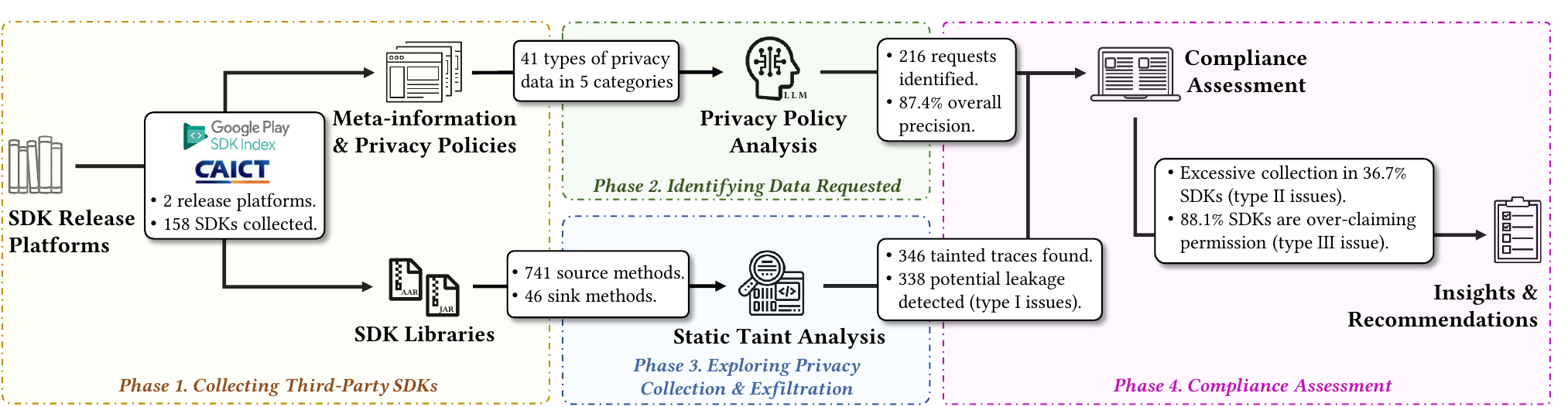}
\vspace{-0.1cm}
\caption{The workflow of potential data exfiltration exploration and compliance assessment at the SDK level}
\label{fig:overview}
\vspace{-0.1cm}
\end{figure*}

We propose a four-phase approach and briefly illustrate it in Figure~\ref{fig:overview}.
In \emph{\textbf{Phase 1}}, we collect 158 Android SDKs, including their binary files, privacy policies, and metadata from two release platforms.
After that, \emph{\textbf{Phase 2}} involves analyzing the privacy policies using state-of-the-art LLMs to identify the types of privacy data requested by the SDKs.
In \emph{\textbf{Phase 3}}, we perform static taint analysis on the SDKs to examine their data collection behaviors at the implementation level, aiming to identify traces that constitute a complete privacy data reading and sharing process. SDKs caught to read and share user privacy would be prone to privacy leakage (i.e., type I issues).
Finally, in \emph{\textbf{Phase 4}}, we assess privacy compliance by cross-checking the data requested in the privacy policies with the actual data collection behaviors, aiming to detect excessive collection (Type II) and over-claiming (Type III) issues.

\begin{comment}

We propose a four-phase approach and briefly illustrate it in Figure~\ref{fig:overview}.
Our study begins with collecting as many Android SDKs as possible from maistream release platforms, shown as \emph{\textbf{Phase 1}}.
As a result, we managed to collect 158 SDKs, including their binary files, the privacy policies, and meta information from the produce page.
After that, we analyze the privacy policies of the collected SDKs, from which we identify data requested by the SDKs (\emph{\textbf{Phase 2}}). 
Specifically, we resort to state-of-the-art LLMs to analyze what types of privacy data are requested in natural language.
On the other hand, we explore data collection behaviors and potential data exfiltration of the collected SDK at the implementation level (\emph{\textbf{Phase 3}}).
To this end, we perform an automatic static taint analysis for the collected SDKs, aiming to identify the traces in the SDK implementation that constitute a complete privacy data reading and sharing process. 
SDKs attempts to read and share privacy data will be assessed to be prone to privacy leakage (i.e., type I issues).
In the end, we assess the \emph{status quo} of privacy compliance by cross-checking the data requested in the privacy policy and the data collection behaviors of each collected SDK (\emph{\textbf{Phase 4}}). Through this assessment, we aim to identify the excessive personal data access and permission over-claiming issues (type II and type III) from the collected SDKs.
\end{comment}

% !TeX root = ../main.tex

%\section{Scope of Privacy Data}
\subsection{Scope of Privacy Data}
\label{sec:scope}

Considering most privacy data stipulates an explicit permission for apps to access, we mainly resort to the list of permissions~\cite{google2024manifest} from Android developer documentation to define the list of privacy data. 
%Specifically, we browse the latest list of permissions defined in Android OS~\cite{google2024manifest} to retrieve the data that request explicit permissions to access, and from which we identify privacy data, such as users' location, contacts, and calendars. 
The list is further complemented with privacy-sensitive items recognized from the \emph{privacy or security change} section of each Android OS release report, for example, the clipboard and a group of user-unresettable identifiers (UUIs) are recognized in the release document of Android 10~\cite{google2023privacy10} and added to the scope of this study. 
% We resort to the Android developer documentation, including the \hilight{OS release report and API documents}, \baigd{too general. Maybe just the permissions and API documents?, given that some permissions are meant to protect data. Give citations as well}to define the list of data considered as user privacy. By doing this, most user-unresettable identifiers and privacy-sensitive items (e.g., account information) can be recognized since they have been emphasized in the privacy changes of Android OS upgrade~\cite{google2023changes8,google2023privacy10}. \baigd{I don't see a clear link between change and data types}
In addition, we also take existing literature in personal information collection~\cite{chen2014information,leung2016should,papadopoulos2017long,ren2018longitudinal,ren2016recon,wang2021understanding} and app analysis~\cite{enck2014taintdroid,gibler2012androidleaks,qiu2015apptrace,sun2016taintart} into consideration to complement the list of privacy data. 
We thus include the usage of diverse sensors (e.g., motion and light sensors) into the list because they have been widely studied in the research community.
As a result, we recognize 41 types of privacy data and categorize them into five groups.
These 41 data types constitute $\mathbb{D}$ defined in Section~\ref{sec:problem_definitions} and formulate the scope of our compliance assessment.
The list of recognized privacy data is shown in Table~\ref{tab:table-source-stat}.

% !TeX root = ../../main.tex
\begin{table}[t]
\centering
% Config the row spacing
\def\arraystretch{1.05}
% Config the column spacing
\setlength{\tabcolsep}{1pt}
\caption{\label{tab:table-source-stat} List of 41 pre-identified privacy data types and the distribution of the 741 source methods}
\footnotesize
\vspace{-0.1cm}
\resizebox{.48\textwidth}{!}{%
\begin{tabular}{lr|H|lr}
\hline
\textbf{Data type \& class}                   & \textbf{Count}     &  & \textbf{Data type \& class}                  & \textbf{Count} \\ \hline
\multicolumn{2}{l|}{\textbf{(C1) Chip, Cellular \& Peripheral\qquad\qquad}} &  & \,\,Wifi                                & 38             \\
\,\,Call phone                                     & 41     &  & \,\,Wifi MACs                                           & 5              \\
\,\,Carrier info                                   & 5      &  & \,\,\textit{Subtotal}                                   & \textit{(245)}          \\ \cline{4-5} 
\,\,Cellbroadcast                                  & 2      &  & \multicolumn{2}{l}{\textbf{(C3) Location \& Sensors}}\\
\,\,External storage                               & 18     &  & \,\,Camera                                              & 54             \\
\,\,ICCID                                          & 15     &  & \,\,Location\textsuperscript{\textdagger}                             & 87             \\
\,\,IMEI                                           & 5      &  & \,\,Misc. sensors                                       & 140            \\
\,\,IMSI                                           & 3      &  & \,\,\textit{Subtotal}                                   & \textit{(281)}          \\ \cline{4-5} 
\,\,MEID                                           & 2      &  & \multicolumn{2}{l}{\textbf{(C4) Media \& Software Specific\qquad\qquad}}\\
\,\,Network info                                   & 6      &  & \,\,Android ID                                          & 2              \\
\,\,Phone status                                   & 1      &  & \,\,App list                                            & 7              \\
\,\,SD card serial                                 & 4      &  & \,\,Audio record                                        & 22             \\
\,\,Serial                                         & 4      &  & \,\,Google Ad ID        & 3              \\
\,\,SIM info                                       & 7      &  & \,\,Media location                                      & 1              \\
\,\,SMS                                            & 23     &  & \,\,OAID               & 9              \\
\,\,Telephone number                               & 12     &  & \,\,Screen record                                       & 1              \\ 
\,\,\textit{Subtotal}                    & \textit{(148)}   &  & \,\,\textit{Subtotal}                                   & \textit{(45)}\\ \cline{1-2}\cline{4-5} 
\multicolumn{2}{l|}{\textbf{(C2) Wireless Communication}}   &  & \multicolumn{2}{l}{\textbf{(C5) Personal Data}}              \\
\,\,Bluetooth                                      & 117    &  & \,\,Account info                                        & 6              \\
\,\,Bluetooth call                                 & 6      &  & \,\,Browser bookmarks                                   & 1              \\
\,\,Bluetooth ID                                   & 2      &  & \,\,Calendar                                            & 2              \\
\,\,Bluetooth MAC                                  & 36     &  & \,\,Clipboard data                                      & 2              \\
\,\,IP address                                     & 2      &  & \,\,Contact list                                        & 9              \\
\,\,SIP service (VOIP)                             & 21     &  & \,\,Contact log                                         & 2              \\
\,\,SSID/BSSID                                     & 17     &  & \,\,\textit{Subtotal}                                   & \textit{(22)}           \\ \cline{4-5} 
\,\,Ultra-wideband                                 & 1      &  & \textbf{Total}                                          & \textbf{741}  \\\hline
\end{tabular}
}
\begin{flushleft}
\begin{footnotesize}
\textsuperscript{\textdagger} Includes coarse location that is determined by network, and fine location that is jointly determined by network and GPS.\\
\end{footnotesize}	
\end{flushleft}

\vspace{-0.3cm}
\end{table}

\paragraph{Categorization} We then categorize the identified 41 types of privacy data by three criteria:

\begin{itemize}
	\item We primarily group those privacy data that share a similar functionality or purpose. This is to ease our privacy policy based compliance assessment, because many developers may tend to only declare the data to be collected roughly by functionality (e.g., device identifiers, location, etc.) rather than accessing specific data or certain APIs.
	\item We then refine our categorization by grouping those data that can be programmatically accessed through APIs in the same or similar classes based on the Android developer's documentation. This is to facilitate our next step's taint analysis as we need to collect as many involved APIs as possible to be the source methods, which we will detail in Section~\ref{sec:taintanalysis}. 
	\item We also consider the API permissions of collecting those privacy data into our categorization. Due to the historic changes in Android privacy mechanism are mainly reflected in permission rules of calling sensitive APIs, grouping personal data that involve APIs mentioned in the same privacy change can facilitate our discussion and gain insights if any non-compliance is observed. A typical example is that the invocation permission of most APIs to read device identifiers including serial, IMEI, and ICCID, are escalated in Android 10. We accordingly group these aforementioned data into one category.
\end{itemize}

Based on the criterion listed above, we further categorize them into five classes, namely \textit{chip, cellular \& peripheral} (\textbf{C1}), \textit{wireless communication} (\textbf{C2}), \textit{camera, location, and other sensors}, (\textbf{C3}), \textit{media and software specific} (\textbf{C4}), and \textit{miscellaneous personal data} (\textbf{C5}). 
%More detailed descriptions of the five categories can be found in the Appendix~\ref{sec:appendix-data-cat}.

\paragraph{Ontology for Semantic Relationships}
Considering the request of personal data in privacy policies may not always written in a precise manner as defined in the developer's documentation, and sometimes certain privacy data items have multiple names or colloquial phrases, we refer to relevant research~\cite{andow2020actions,bui2021consistency} and propose an ontology to capture the semantic relationship among the 41 types of privacy data, and consequently to guide our assessment.

First, we manually collect alternative names or colloquial phrases for each type of identified privacy data and treat them as \emph{synonyms} ($\equiv$).
For example, we find that the advertising ID and GAID are interchangeably used in privacy policy documents to refer to the Google Ad ID. Thus, all the synonymous terms of a privacy data type will be considered in our assessment.

Besides that, we define a \emph{hypernymy relationship} ($\sqsubset$) between privacy data types in generic and specific terms. For example, we learn that both the user name and email are compositions of the account information of an Android device, and accordingly treat both user name and email \emph{hyponyms} of the term account information (denoted as $\texttt{name} \sqsubset \texttt{account\_info}$ and $\texttt{email} \sqsubset \texttt{account\_info}$).
In case an SDK developer claims to request the email address of the device owner, we assume the API for reading account information will be invoked, and therefore, both the email and name of the user will be accessed.
% Similarly, ``device identifiers'' in the context of Android OS may refer to different technical terms (e.g., IMEI or MEID) according to the compatible cellular network. For that reason, we define both IMEI and MEID as the hyponyms of the term device identifiers.

We also learned that some SDK developers use the rough term ``device identifiers'' in disclaiming privacy data to collect. Unlike the definition of ``\texttt{DeviceID}'' in Android documentation, device identifiers are often referred to as a broad sense of hardware identifiers in privacy policies, such as the serial number of the smartphone. 
Thus, we consider device identifiers as a hypernym of diverse hardware identifiers (e.g., $\texttt{serial} \sqsubset \texttt{device\_identifiers}$). We hold a conservative stance in this study and, thus, assume the involved SDKs claim to request all types of privacy data that fall into the scope of device identifiers, including IMEI, MEID, ICCID, and serial.
We also remark that although a user's location can also be inferred through his/her devices' IP addresses, we assume the term ``location'' and its synonyms indicate the data is derived from SDK's access to location services of Android OS (e.g., \texttt{LocationManager}). The main reason is that we believe the SDK developers are supposed to use precise and unambiguous language in writing their privacy policies. 
Besides that, inferring users' location through different OS components involves different privacy mechanisms and requests for different permissions. 
Thus, an SDK's claiming to access users' locations through any manners other than location services without an explicit explanation would be treated as non-compliance in this study.

\subsection{Phase 1: Collecting Android SDKs}
\label{sec:sdkcollection}

Our study starts by collecting as many Android SDKs as possible from mainstream sources. Since our approach involves cross-checking SDK privacy disclosures with actual data collection behaviors at the implementation level, we gather not only SDK binaries (typically in AAR and JAR formats) but also the privacy policies and relevant metadata provided by the developers, when available.

\paragraph{Sourcing SDKs from Multiple Repositories}
We identify the Google Play SDK Index as the primary source for collecting Android SDKs. Additionally, considering the uniqueness of mainland China's Android ecosystem where Google Play infrastructure is absent, we use CAICT as a supplementary source. 
As a result, we download the latest version of \allsdk distinct SDKs as of October 2022, among which \globalsdk are from the Google Play SDK Index and \chinasdk are released on the CAICT website. 
These \allsdk constitute our previously defined $\mathbb{S}$ (Section~\ref{sec:problem_definitions}) for the later assessment. We also revisit the latest versions of the same group of SDKs in October 2023 and conduct an additional assessment, which we will detail in Section~\ref{sec:longitudinal}.

We remark that, at the time of this study, both the Google Play SDK Index and CAICT have not enforced strict regulation or code auditing mechanisms like the Google Play App Store. As a result, there is no guarantee that the SDKs from these sources include comprehensive documentation or privacy policies required by regulations such as GDPR.

\paragraph{Crawling Meta Information and Privacy Policies}
Next, we implement a crawler with Selenium~\cite{selenium} to retrieve meta information and privacy policy links from the two sources. 
The meta information includes the SDK name, provider (developer's identity), Maven IDs, and functional categories. Such information will be jointly used to uniquely identify SDKs during our subsequent analysis.
We then collect the privacy policies from the retrieved links. We will use them to extract the list of personal data that the SDK requests.

The retrieval of privacy policies is found more complex than collecting meta information due to the lack of a compulsory standard for drafting them. For SDKs from the Google Play SDK Index, we rely on the links provided in the ``data safety section'' section if available, and download the corresponding webpages. Similarly, for SDKs from the CAICT website, we download the webpages from the ``SDK privacy policy'' link on the SDK details page. We use Google Translate to process all non-English text on these webpages.

Due to the limited availability of privacy policies, we only collected policies for 52 SDKs directly from the two platforms. For those do not provide privacy policy URLs, we manually searched the Internet using keywords like ``disclaimer'', ``policy'', ``disclosure'', and ``guidance'' to look for any available privacy policies drafted by the SDKs' developers. As a result, we identify privacy policy webpages for another 57 SDKs. 
In total, we obtained valid privacy policies for 109 SDKs (69.0\% of all SDKs). The text from these webpages will be extracted for further analysis.

\subsection{Phase 2: Identifying Data Requested in Privacy Policies}
\label{sec:pp-analysis}

After collecting the 109 privacy policies, our next step aims to identify the data requested by each SDK from them.
In contrast to traditional app privacy policies that are written in the \emph{user-oriented} language, SDK privacy policies are more \emph{developer-oriented} and may contain many domain-specific terminology (e.g., ICCID, BSSID) and technical synonyms (e.g., Google Ad ID, which also appears as Advertising ID, or GAID). 
This uniqueness makes it challenging for existing methods~\cite{andow2019policylint,andow2020actions,bui2021consistency,xie2022scrutinizing,yu2016can,zimmeck2016automated} to extract domain-specific entities accurately.
To tackle these challenges, we leverage an LLM-based  Natural Language Inference (NLI) model to infer the data collection practice from the privacy policies.

\begin{figure}[t]
\centering
% trim order: <left> <lower> <right> <upper>}
\includegraphics[trim=0.2cm 6.1cm 0.2cm 0cm,clip=true,width=1\linewidth]{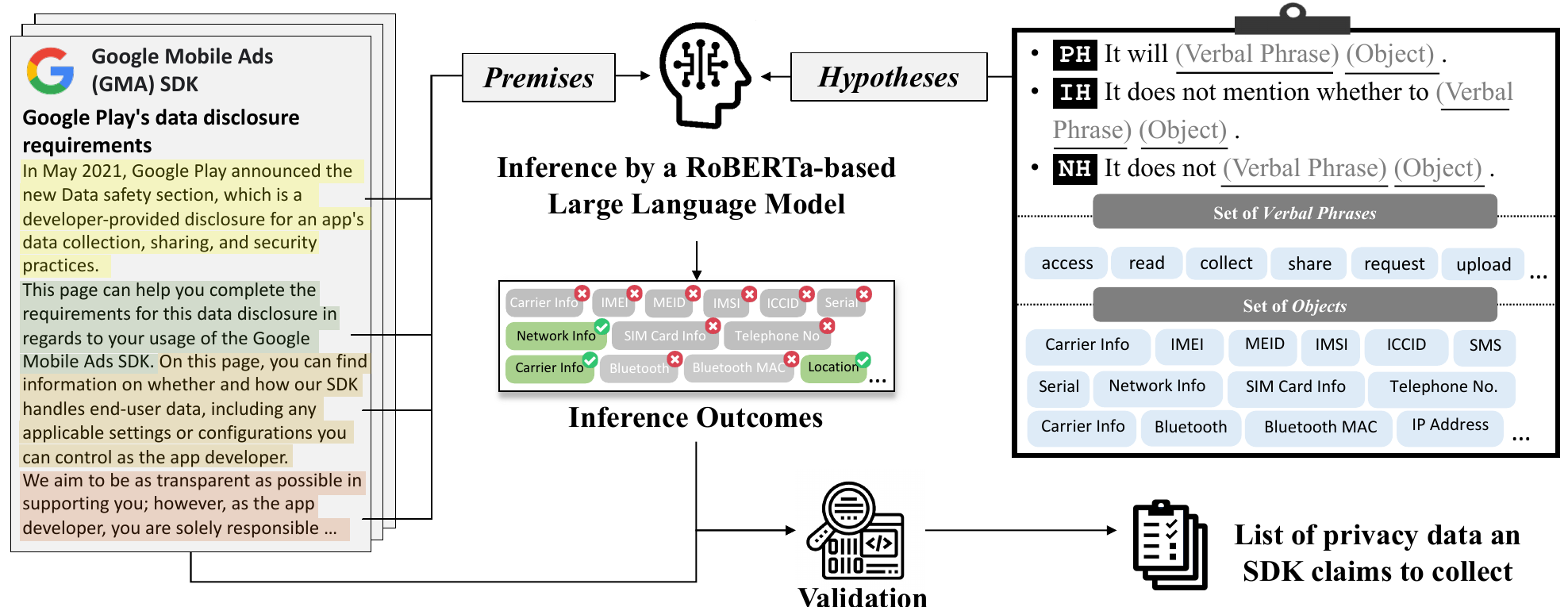}
\vspace{-0.2cm}
\caption{The workflow of our privacy policy analysis to find out what an SDK claims to collect}
\label{fig:llm-inference}
\vspace{-0.1cm}
\end{figure}

\paragraph{Adoption of NLI Model}
NLI has proven effective in understanding natural language across different domains~\cite{harkous2022hark}. 
In this work, we use the state-of-the-art LLM-based NLI model named \texttt{roberta-NLI}\footnote{The full name of the model is \texttt{roberta-large-snli\_mnli\_fever\_anli} \texttt{\_R1\_R2\_R3-nli}. We shortly write as \texttt{roberta-NLI} to save space.} to identify requested data in privacy policies. 
By parsing sentences from these policies as \textit{premises}, we input them into the model alongside \textit{hypotheses} based on 41 pre-identified data types (refer to Section~\ref{sec:scope}), allowing the model to assess the entailment relationship between them.
We briefly illustrate our approach in Figure~\ref{fig:llm-inference}. 

There are two key challenges arose during the hypothesis formulation. 
The first challenge is the missing of a canonical writing standard of SDK privacy policies.
Privacy policies often use developer-oriented language with technical synonyms, requiring us to manually gather synonyms (e.g., the advertising ID and GAID are interchangeably used to refer to the Google Ad ID) to ensure hypothesis completeness. 
Besides that, we also tailored hypotheses by matching different verb phases (e.g., ``collect'', ``read'') to each data type and its synonyms to maximize the entailment score when the collection of the inferred data type is mentioned in the privacy policy.

The second challenge is handling the complex context of SDK privacy policies, as the NLI model is not domain-specifically trained. 
For example, we found common sentence formats in privacy policies, such as ``According to GDPR, all collection of device identifiers must be declared in advance.'' 
Conventional inference models might incorrectly conclude with high confidence that this implies the \textit{collection of device identifiers}, but it is irrelevant to the actual behaviors of the SDK being analyzed.
To address this issue, we created three tailored hypotheses for each combination of verbal phrase and object, namely the \emph{positive hypothesis} (shown as \textbf{PH} in Figure~\ref{fig:llm-inference}, e.g., ``It will collect ...''), \emph{negative hypothesis} (\textbf{NH}, e.g., ``It does not collect ...''), and \emph{irrelevant hypothesis} (\textbf{IR}, e.g., ``It does not mention whether to collect ...''). 
As a result, we have composed a list of 507 hypotheses covering the 41 pre-identified data types.%\footnote{The full list of hypotheses is provided in Appendix~\ref{sec:appendix-hypotheses}.}

\paragraph{Inference of Data Requested}
Given three hypotheses have been defined for each combination of verbal phrase and object, we only consider a data type claimed to be collected by an SDK if there exists \emph{at least one} combination of the associated object (e.g., location) and applicable verbal phrase (e.g., request) satisfying the logical condition as follows:

\vspace{-0.35cm}
\[(\text{Score}_{p}> T) \wedge (\text{Score}_{p} > \text{Score}_{n}) \wedge  (\text{Score}_{p} > \text{Score}_{i}) \]

where $\text{Score}_{p}$ indicates the entailment confidence score for the \emph{positive hypothesis}. An entailment inference requires $\text{Score}_{p}$ to be greater than a threshold value, written as $T$. Simultaneously, $\text{Score}_{p}$ must be greater than the scores for \emph{negative} and \emph{irrelevant hypotheses} (i.e., $\text{Score}_{n}$ and $\text{Score}_{i}$), indicating that the premise does not only ``mention'' the collection of that data type but also confirm that collection behavior is ``performed'' by the SDK itself. 

\paragraph{Finding the Optimal Threshold}
Intuitively, to infer a specific type of data being collected, the entailment prediction score ($\text{Score}_{p}$) is supposed to be greater than 0.5 to ensure it is always higher than the prediction of the neutral and contradiction hypotheses.
However, our preliminary experiments observe a large number of false positives by setting the threshold to 0.5. 
For that reason, we perform a small-scale analysis to tune the threshold value from 0.5 and above until we find an optimal value that minimizes the number of false positive predictions.
Specifically, we assume that Google should provide high-quality privacy policies for their own SDKs, and therefore we collected five distinct privacy policy shared by 13 Google in-house SDKs. 
We manually analyzed declared privacy data types from the five privacy policies to set a ground truth.
We then fine-tuned the threshold value and discovered that the NLI model could achieve perfect prediction with a threshold of 0.73.
For that reason, we adopt 0.73 as the optimal threshold in our large-scale privacy policy analysis. 
% !TeX root = ../main.tex
\subsection{Phase 3: Exploring Privacy Collection and Leakages}
\label{sec:taintanalysis}

We leverage the static taint analysis techniques to explore privacy collection behaviors and potential exfiltration in SDKs.
We define a valid privacy data exfiltration is constituted of a \textit{collection action} that queries one or more privacy data from the OS, and a \textit{sharing action} that shares the obtained privacy data, directly or indirectly, to the public spaces including the Internet, system settings, and the public storage.
Thus, we consider that an SDK engages in privacy collection if its implementation contains a data flow originating from a privacy data collection action. This privacy collection leads to exfiltration if the data flow ends at a sensitive sharing action.

\begin{comment}
We assume that a valid privacy leakage is constituted of an \textit{collection action} that queries one or more privacy data from the OS, and a \textit{sharing action} that the tested SDK shares the obtained privacy data, directly or indirectly, to the public spaces including the Internet, system settings, and the public storage of the device.

From previous steps (i.e., Phase 1), we find that not all APIs to read our 41 pre-identified privacy data are interface methods that are exclusively designed for reading certain data, for example, the API \texttt{getDeviceId()} in a class named \texttt{TelephonyManager} is merely designed to read the device ID. Instead, some APIs are provided as class attributes (e.g., one API to read serial number in \texttt{android.os.Build} class is declared as a public string constant) or general-purpose interfaces (e.g., read through system settings).
For those reasons, we leverage an open-source static taint analysis toolkit named \emph{Appshark}~\cite{appshark}.
Appshark supports configuring both methods and variables as the source, and therefore is an ideal tool for us in this study.
On the basis of the underlying Appshark framework, we disable the functionality that is irrelevant to our study to improve the analysis efficiency. 
We also customize our taint analysis to only trace the collection and sharing actions within the package scope of the embedding SDKs, aiming to further speed up the analysis.
	
\end{comment}

\paragraph{Configuring Taint Analysis}
We aim to collect all relevant APIs that contribute to a collection action and include them into the \textit{source methods} for taint analysis.
To this end, We refer to an existing research~\cite{rasthofer2014machine} which proposes a technique named \texttt{SuSi}, which classifies over 900 API methods that can potentially leak users' privacy in Android 4.3 and earlier versions. 
Considering there have been multiple version's evolutions since Android 4.3 that the API rules may have experienced significant change, we first map the APIs identified in~\cite{rasthofer2014machine} with the 41 types of privacy data. We iterate the developers' documents of those APIs and filter out those APIs that are no longer valid in Android Open Source Project (AOSP) implementation. We then manually visit the developer's document and AOSP source code to collect as many APIs that are associated with the new data types as possible.
As a result, we collect 741 source methods.\footnote{We count the number of methods by unique method signatures in this paper.}
%To facilitate our analysis and discussion, we further categorize the recognized privacy data into five classes.
The distribution of 741 methods corresponding to the 41 types of privacy data can be found in Table~\ref{tab:table-source-stat}.

% !TeX root = ../../main.tex

\begin{table}[t]
%\begin{wraptable}{r}{6.7cm} % only used for single-column environment
\centering
% Config the row spacing
\def\arraystretch{1.1}
% Config the column spacing
\setlength{\tabcolsep}{3pt}
\caption{\label{tab:table-sink-stat} A brief statistics of recognized sink methods}
\footnotesize
\scriptsize
\vspace{-0.1cm}
%\begin{footnotesize}
%\centering
\begin{tabular}{llr}
\hline
\textbf{Purpose}                  & \textbf{Class name}                               & \textbf{\makecell[lt]{Count}} \\ \hline
Network Transfer\,\,\,\,          & \texttt{java.net.HttpURLConnection}               & 1              \\ \cline{2-3} 
                                  & \texttt{java.net.URL}                             & 2              \\ \cline{2-3} 
                                  & \texttt{java.net.URLConnection}                   & 2              \\ \cline{2-3} 
                                  & \texttt{org.android.spdy.SpdyRequest}             & 12             \\ \cline{2-3} 
                                  & \texttt{org.android.spdy.SpdySession}             & 3              \\ \cline{2-3} 
                                  & \texttt{org.apache.http.client.methods.HttpGet}   & 8              \\ \cline{2-3} 
                                  & \texttt{org.apache.http.client.methods.HttpPost}  & 8              \\ \cline{2-3} 
                                  & \texttt{org.apache.http.params.HttpParams}        & 1              \\ \hline
File Saving                        & \texttt{java.io.Writer}                           & 5              \\ \cline{2-3} 
                                  & \texttt{java.io.FileOutputStream}                 & 1              \\ \hline
System Settings                   & \texttt{android.provider.Settings}                & 3              \\ \hline
\textbf{Total}                    &                                                   & \textbf{46}    \\ \hline
                                  
\end{tabular}
%\end{footnotesize}

\end{table}
% \end{wraptable} % only used for single-column environment

Next, we retrieve all methods that perform a sharing action from Android SDKs. Although there are not many new APIs introduced after Android 4.2 that can be treated as sink methods, we find most sink methods studied in earlier literature including  \texttt{SuSi}~\cite{rasthofer2014machine}, such as logging and SQLite databases, are no longer valid in the latest Android versions and therefore be excluded in this study~\cite{google2023loginfo,google2022sandbox}.
%For example, logging is no longer a feasible approach for sharing data among different apps because an app cannot read logs created by another~\cite{google2023loginfo}. Similarly, methods to access and manage SQLite databases are not counted because the database files are protected by the sandbox mechanism of Android and are not visible to other apps~\cite{google2022sandbox}.
In the end, we identify 46 methods from 11 classes and then configure them as \textit{sink methods}. 
We present a brief statistics of the sink methods by their containing classes in Table~\ref{tab:table-sink-stat}.

The majority of sink methods are relevant to data transfer via network interfaces, reserving 37 out of 46 identified method signatures.
Besides that, six method are recognized because they can share sensitive data via public storage. The remaining three sink methods are counted because they can be used for modifying the system settings, which is known as an undocumented channel to share data among different parties on the device~\cite{meng2023post}. We note that writing contents into the system settings requires privileged permission and we will discuss it in Section~\ref{sec:issues}.

% We also remark that we exclude some sink methods that are widely used in early Android versions or other platforms. For example, the logging mechanism is not a feasible approach for sharing data among different apps because an app cannot read logs created by another~\cite{google2023loginfo}. Similarly, methods to access and manage SQLite databases are not counted because the database files are protected by the sandbox mechanism of Android and are not visible to other apps~\cite{google2022sandbox}. 

% \input{sections/compliance-assessment}
%% \input{sections/policy-analysis}
% End of components of the methodology section

% !TeX root = ../main.tex
\section{Evaluation}
\label{sec:evaluation}

Our evaluation aims to assess the performance of our proposed NLI model-based privacy policy analysis and explore the privacy data collection and potential leakages in the SDKs. 
After that, we analyze the privacy preservation landscape at the SDK level by cross-checking the data requested in the privacy policies and the data collected in SDKs' implementation (refer to Phase 4 of Figure~\ref{fig:overview}).

\paragraph{Implementation}
We implement our privacy policy analysis in Python with PyTorch. The NLI model is provided by a Python package named \emph{Transformers}.\footnote{Available at \url{https://huggingface.co/docs/transformers/index}}

\paragraph{Adoption of Automatic Taint Analysis}
We learned that there are some off-the-shelf automatic static taint analysis tools for Android apps.
Our preliminary investigation considers three open-sourced tools, namely FlowDroid~\cite{arzt2014flowdroid}, Mariana Trench~\cite{mariana_trench}, and AppShark~\cite{appshark}.
We set up these three tools on our experimental PC, feed four best-selling non-game apps from the Google Play App Store, namely Facebook, Instagram, Snapchat, and TikTok, into these tools, and explore data flows simultaneously containing a privacy collection action and a sharing action.
Our observation shows that AppShark performs the best in terms of time efficiency, success rate of analysis, and capability of handling inter-process communication (IPC).
For those reasons, we adopt AppShark for the taint analysis of SDKs on a large scale. 
%For the adoption of automatic taint analysis, we carried out a pilot study to find an off-the-shelf tool that best suited our study, and we consequently selected AppShark~\cite{appshark} for our taint analysis. We detail our pilot study in Appendix~\ref{sec:appendix-pilot-study-taint-analysis-tool} to save space. 
%Our static taint analysis is implemented in Java.% and it supports users to specify target apps and customize rules for taint analysis in a configuration file.
For each SDK, we implement a blank Android app to embed it and pass the app package to the automatic taint analysis program.
% Considering that taint analysis may unavoidably produce false positives, manual validation of the tainted traces is performed to ensure the correctness of our further assessment.
%\baigd{Will this introduce unreachable code?}
%We plan to open-source our implementation and relevant artifacts in the future.
% We will open-source our implementation and relevant artifacts upon paper acceptance.

\paragraph{Evaluation Goals}
Our evaluation aims to answer the following three research questions (RQs):

\begin{itemize}
\item \textbf{RQ1}. Can our privacy policy analysis identify the privacy data claimed to be collected by SDKs on a large scale? What is the performance of our NLI model-based approach?

\item \textbf{RQ2}. Can our taint analysis find privacy data collection and sharing behaviors from the SDKs? 
How many tainted traces are detected? % What type of data are mostly collected and/or shared? 
Can those tainted traces (really) leak users' privacy? 
%Have our detected potential privacy leakages (already) been exploited in existing SDKs?  

\item \textbf{RQ3}. By cross-checking their privacy policies, what is the \textit{status quo} of privacy data protection at the SDK level? Do the collected SDKs comply with their privacy policies?
\end{itemize}

% !TeX root = ../../main.tex

%\subsection{RQ1: Privacy Policy Analysis}
\subsection{RQ1: Privacy Policy Analysis Performance}
\label{sec:rq1}

\begin{table*}[t]
\centering
\caption{\label{tab:table-pp-analysis-stat} Evaluation outcomes of privacy policy analysis (only data types requested are displayed)}
\vspace{-0.15cm}
% trim order: <left> <lower> <right> <upper>}
\includegraphics[trim=0cm 0cm 0cm 0cm,clip=true,width=0.9\linewidth]{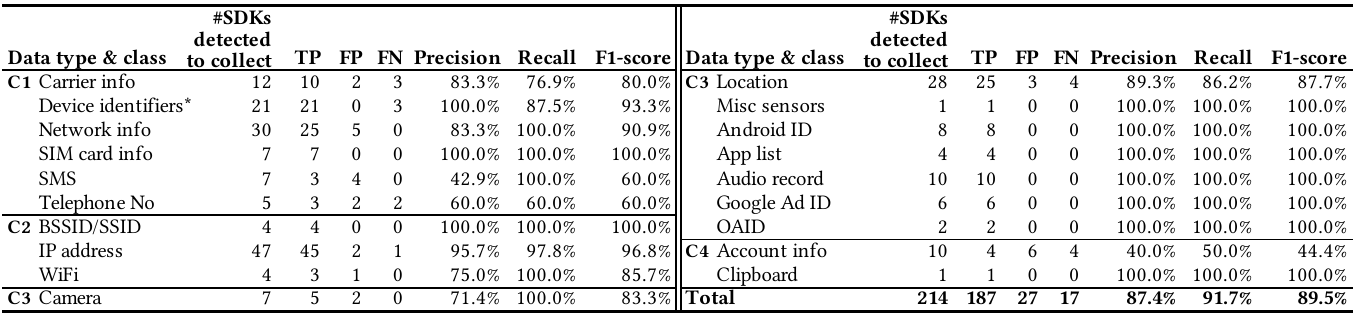}
\vspace{-0.2cm}
\begin{flushleft}\footnotesize\smallskip
* We use a broad sense of device identifiers to represent all hardware UUIs, including IMEI, MEID, IMSI, ICCID, and serial.
\end{flushleft}\vspace{-0.2cm}
\end{table*}

We perform NLI model-based analysis on the 109 SDK privacy policies and present the analysis outcomes in Table~\ref{tab:table-pp-analysis-stat}.
Our analysis identifies \totalrequestinppbymodel data requests from privacy policies. 
Among the \totalrequestinppbymodel data requests, IP address reserves the largest share, contributing to 47 (21.9\%) occurrences. network information and location are the remaining data types among the top three.
From the perspective of SDKs, the number of data types claimed by an SDK ranges from 0 to 6, with the average number 1.9.

% [USENIX submission] Among the \totalrequestinpp data requests, the network information reserves the largest share, contributing to 63 (13.5\%) occurrences. IP address and account information are the remaining two data types among the top three, which have been requested by 49 (10.6\%) and 47 (10.1\%) occurrences, respectively. 
% [NDSS submission] Among the 464 data requests, the device identifiers in a broad sense (including IMEI, MEID, serial, etc.) reserve the largest share, contributing to 59 occurrences. Account information and network information are the remaining two data types among the top three, which have been requested by 47 and 38 occurrences, respectively.

We then conduct a manual inspection of our findings in the 109 privacy policies. As a result, we find \totalrequestinpp verified data requests covering 19 data types, and accordingly treat these requests as the ground truth in evaluating our analysis approach.
The results demonstrate that our approach can effectively analyze privacy policies on a large scale, achieving a promising precision of 87.4\% and an F1 score of 89.5\%.
(see Table~\ref{tab:table-pp-analysis-stat}).

We find our NLI model-based analysis achieves perfect precision in detecting data collection for 10 data types despite 27 \emph{false positive} cases being observed, i.e., wrongly recognized by the NLI model. 
Our evaluation outcomes show that account information (6), network information (5), and SMS (4) are the top three data types involved in false positives, indicating the model struggled to infer data collection for certain types.
These errors likely stem from the model's misinterpretation of terms and data type scopes. 
For example, although our NLI model correctly recognizes SMS, it fails to differentiate SMS and other messaging methods like emails, leading to high-confidence false positives while dealing with sentences like ``we may send email to the users...'' as SMS collection.
\subsection{RQ2: Privacy Collection and Potential Leakages at the SDK Level}
\label{sec:rq2}

Our static taint analysis identified \accessallsdk instances of privacy data access across 27 of the 41 pre-identified data types. Out of \allsdk SDKs, 95 (60.1\%) were found to read privacy data, with 34 (21.5\%) having at least one tainted trace. The analysis revealed that \taintallsdk (\taintratioallsdk) of these accesses were tainted, suggesting potential privacy leakage into public spaces.
We present the taint analysis outcomes by data types in Table~\ref{tab:table-taint-analysis-stat}.
% We also performed manual validation to exclude false positive tainted traces, which we will elaborate on with examples in the remainder of this section. 

\begin{table}[t]
\centering
\caption{\label{tab:table-taint-analysis-stat} A brief statistics of the taint analysis results (only data types called and/or tainted are displayed)}
\vspace{-0.2cm}
% trim order: <left> <lower> <right> <upper>}
%\includegraphics[trim=0.0cm 18.5cm 2.4cm 0.2cm,clip=true,width=0.75\linewidth]{images/taint_statistic.pdf}
\includegraphics[trim=0.0cm 0cm 0cm 0cm,clip=true,width=1\linewidth]{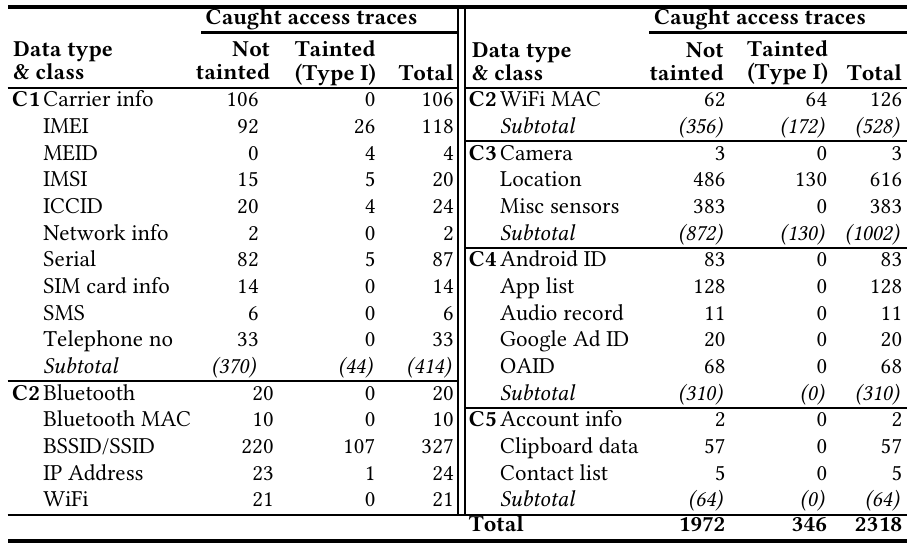}
\vspace{-0.2cm}
\end{table}

\paragraph{Distribution of Data Collected \& Shared}
As shown in Table~\ref{tab:table-taint-analysis-stat}, location and sensors' data (category C3) reserves the greatest portion of collection actions, contributing 1002 out of 2318 (43.2\%) caught collection actions.
The remaining privacy data categories ordered by their number of caught collection actions are data for wireless communication (C2, 22.8\%), data for chip, cellular and peripheral (C1, 17.9\%), media and software-specific data (C4, 13.4\%), and personal data (C5, 2.8\%), respectively.

From the privacy data sharing perspective, the access of wireless communication data (C2) constitutes the largest part of the tainted flow, as 172 out of 346 (49.7\%) traces are tainted. The numbers of sharing actions involving data in categories C3 and C1 are following up, reserving 37.6\% and 12.7\% of all sharing actions, respectively. No sharing actions about data in categories C4 and C5 are observed. 
From the \emph{identifiability} perspective, we find that most privacy data involved in sharing actions are referred to as \emph{personally identifiable information} (PII) in relevant literature~\cite{ren2018longitudinal,ren2016recon,shen2021understanding,wang2020exploring}.
%Most privacy data involved in sharing actions are either UUIs or PIIs, according to our categorization of privacy data in Section~\ref{sec:scope}.
% Given that Google has escalated permission requirements or even prohibited apps' access to them through APIs, the caught tainted traces do not automatically imply the actual privacy leakages. We will discuss the feasible leakages later in this section.

We remark that whether SDKs can collect privacy data eventually depends on the permission and privilege of their embedding apps, and tainted traces do not automatically imply the exfiltration in runtime. 
A manual validation is performed to exclude false positive tainted traces. We will elaborate more on them in the remainder of this section.

\begin{comment}    
\begin{framed}
\noindent \textbf{Findings in RQ1:}
\toolname is capable of detecting personal identifier collection from SDKs.
Our study finds that personal privacy access is pervasive in publicly available SDKs, and 21.7\% analyzed SDKs contain at least one trace tainted.
The location and sensor data are the most collected privacy, as 1138 collection actions observed from \allsdk collected SDKs.
\toolname also detects \taintallsdk sharing actions that potentially leak users' privacy to the public. The majority of data involved in the sharing actions are for wireless communication purposes.
\vspace{-0.1cm}
\end{framed}
\end{comment}

% !TeX root = ../../main.tex
%% This tex file continues from the first half of RQ2

% !TeX root = ../../main.tex
\begin{figure}
%\begin{wrapfigure}{r}{0.46\textwidth}  % only used for single-column environment
\begin{lstlisting}[language=Java]
public static String a(Context context) {
  ConnectivityManager cm = (ConnectivityManager) context.getSystemService(Context.CONNECTIVITY_SERVICE);
  NetworkInfo info = cm.getActiveNetworkInfo();
  if (info.isConnected()) { return "wifi"; }
  ...
  try {
    return info == null? "" : info.getTypeName()+"-"+info.getSubtypeName()+"-"+info.getExtraInfo().toLowerCase();
  }
  catch(Exception e) { return ""; }
}
\end{lstlisting}
\vspace{-9pt}
\caption{A false positive example of SSID reading (line 7)}
\label{fig:algorithm-false-positive}
\vspace{-6pt}
\end{figure}
%\end{wrapfigure}  % only used for single-column environment

\paragraph{Potential Privacy Data Leakage (Type I Issues)}
During our analysis, we noticed that static taint analysis cannot well determine the execution paths in conditional and branched code.
If any tainted trace is found spanning over an unreachable, dead, or logically infeasible code, we exclude the tainted trace from the statistics of potential leakages. % and mark it as a false positive. (Mark: maybe mentioning "false positive" here is not a good practice as it may gives reviewers a false image that we are doing statistics here and will discuss FP/FN/F1-score). In our taint analysis, there does not exist FN because all our manual validation is performed based on an assumption that Appshark has found all possible traces, and it is impossible to manually iterate all source code and find traces of over 100 SDKs.
In Figure~\ref{fig:algorithm-false-positive}, we provide an example code reverse-engineered from one of our examined SDKs named ``\texttt{mipush}''. We find it attempts to read the SSID of the connected WiFi but is later assessed by us as a false alarm. The SSID reading action (can be realized by the method \texttt{getExtraInfo} (line 7) until Android 10) would only take place if the device is not connecting to WiFi (after line 4), therefore we determine this is a logically infeasible code regarding privacy data access.  
In this study, we manually exclude the false alarms from the \taintallsdk traces and accordingly identify \correcttaintallsdk (\precisionallsdk) logically valid traces, among which each trace indicates an occurrence of privacy leakage issue. 
%We also provide a case study for the false alarms in Appendix~\ref{sec:appendix-taint-case-study} to demonstrate the intrinsic limitation of static taint analysis with more details.

For the sharing operations, we observe that 332 out of 338 (98.2\%) valid traces are tainted at network interfaces. The six remaining traces are tainted at system setting APIs.
Compared with sharing privacy data through the Internet, we remark that writing privacy data in system settings is a more worrying issue because it exploits the privilege of pre-installed apps and is capable of sharing user privacy even without reading any privacy data from the Android OS.
%Apart from sharing privacy data through the Internet, we find writing privacy data in system settings is a more worrying issue, because it exploits the privilege of pre-installed apps and is capable of sharing user privacy even without reading any privacy data from the Android OS.
We will detail this in Section~\ref{sec:issues}.

\begin{comment}
\begin{framed}
\noindent \textbf{Findings in RQ2:}
\toolname demonstrates promising precision (\precisionallsdk) in finding potential privacy leakages from SDK. 
Our findings also unveil a severe issue in system settings injection, which can be exploited by malicious SDK when it is embedded in a pre-installed app or a system app.
Successful exploitation can bypass the privacy mechanism of Android OS and share a \textit{de facto} UUI in the public space for other apps to access without requesting any permission. 
Two SDKs are found to exploit the system settings injection issue.
\vspace{-0.1cm}
\end{framed}
\end{comment}

% !TeX root = ../../main.tex

%\subsection{RQ3: SDK-level Privacy Compliance}
\subsection{RQ3: SDK-Level Privacy Compliance}
\label{sec:rq3}

We resort to cross-checking the privacy policies with the actual data collection behaviors of the collected SDKs to understand the \emph{status quo} of privacy compliance at the SDK level.
We remark our assessment only focuses on the 109 out of 158 (69.0\%) SDKs that provide privacy policies, and the 27 data types that appear in the data collection behaviors observed from the taint analysis.
We present our assessment results in Table~\ref{tab:table-corsscheck}.

\begin{table}[t]
\caption{\label{tab:table-corsscheck} Assessment of privacy compliance by data types}
\vspace{-0.15cm}
% trim order: <left> <lower> <right> <upper>}
\includegraphics[trim=0.5cm 13.45cm 12.3cm 0.3cm,clip=true,width=1\linewidth]{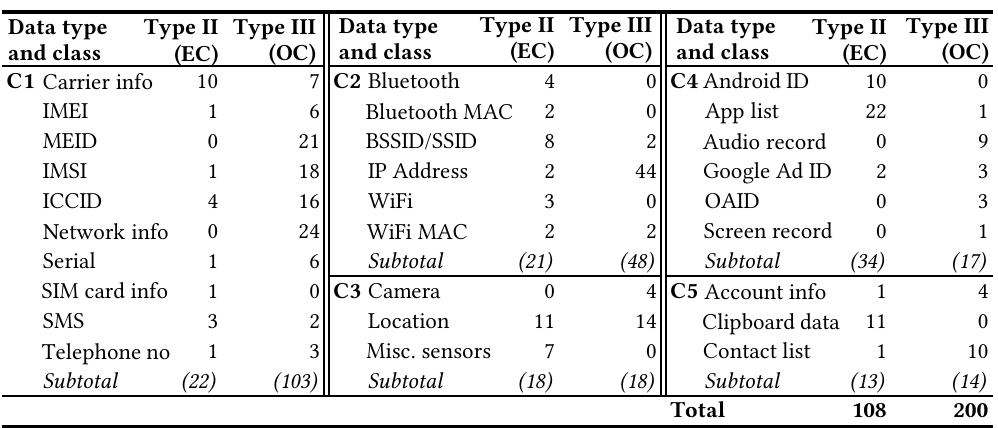}
\vspace{-0.3cm}
\end{table}

\paragraph{Excessive Collection (Type II Issues)}
Our evaluation finds that 40 out of 109 (36.7\%) SDKs collected excessive privacy data compared with their privacy policies (e.g., an SDK collects personal identifiers but does not declare them in its privacy policy). 
Among the 40 SDKs that excessively collect users' privacy, the number of undeclared data types ranges from 1 to 9. Each SDK has an average of three undeclared data types. 
As shown in Table~\ref{tab:table-corsscheck}, 22 out of the 41 pre-identified data types (81.5\%) have been excessively collected by the examined SDKs. 
The app list, clipboard data, and location are the top three most excessively collected data types. 
Specifically, 22 SDKs (20.2\%) collect the app list without any declaration in their privacy policies, followed by clipboard data and location that are excessively collected by 11 (10.0\%) SDKs. 
% We also remark that all caught collection of clipboard data by 11 SDKs are excessive collection without a proper disclosure, by jointly considering our evaluation outcomes in RQ1 and RQ2. (Mark: discuss separated in the next next paragraph)
By jointly considering data collection behaviors (refer to Table~\ref{tab:table-taint-analysis-stat}), our findings show that the app list and clipboard data are the two most \emph{stealthily collected} data types.
Only 4 SDKs collecting the app list have mentioned it in their privacy policies (leading to 22 excessive collection behaviors). None of the 11 caught SDKs have declared to read the clipboard data. 

\paragraph{Over-claiming (Type III Issues)}
Apart from the excessive collection, we find that the over-claiming issue in privacy policies is far more pervasive in the tested SDKs. 
In our study, 96 out of 109 (88.1\%) SDKs are found to claim more data types in privacy policies than what they (really) need. 
The IP address, network information, and MEID are the top three data types that have been over-claimed in the tested SDK privacy policies. 
Especially, 44 out of 109 (40.4\%) SDKs that provide privacy policies over-claim their access to IP addresses.
Such pervasive claims, again, reflect the worrisome landscape of privacy compliance at the SDK level.

\begin{comment}
From the privacy data perspective, our findings show that the app list and clipboard data are the two most \emph{stealthily collected} data types.
None of the 11 caught SDKs have declared to read the clipboard data. Only three out of 25 SDKs collecting the app list have mentioned it in their privacy policies (leading to 22 excessive collection behaviors). 
The IP address, IMEI, MEID, and IMSI are the top four data types that have been over-claimed in SDKs' privacy policy. 
It is worth noting that 75 out of 109 (68.8\%) SDKs that provide privacy policies over-claim their access to IP addresses.
Such pervasive claims, again, reflect the worrisome landscape of privacy compliance at the SDK level.
\end{comment}

\paragraph{Lessons Learned}
Our study reveals two key findings. First, some SDK developers use vague or broad terms like ``identifiers'' or ``device information'' instead of specifying the exact data being collected, leading to differing interpretations. For instance, while most developers use ``identifiers'' to indicate device identifiers, some refer to advertisement identifiers, causing over-claims for device identifiers and under-claims for Android ID or Google Ad ID. Second, although many SDKs declare access to both GPS and IP addresses for location data, our taint analysis shows that most SDKs only use location APIs, leading to over-claiming cases of IP addresses. This lack of clarity can confuse users about what data is actually collected, reducing transparency in privacy policies.

\begin{comment}
\begin{framed}
\noindent \textbf{Findings in RQ3:}
In the latest Android ecosystem, excessive collection of privacy data (36.6\%) and over-claiming collected data types (88.1\%) in privacy policies are prevailing issues. 
The inconsistency between privacy policies and actual practices indicates a poor status quo of SDK-level privacy protection compliance. 
App list, clipboard data, and location are the top three most excessively collected privacy data types. 
The IP address, IMEI, and Google Ad ID are the top three data types that have been over-claimed in the tested SDKs privacy policies. 
% Various personal identifiers are found to be stealthily collected by SDKs including location, carrier info, app list, and Android ID.
The use of ambiguous terminologies and the prevalent absence of privacy policies for SDKs reflects the non-standardization and deficiency of privacy protection at the SDK level. 
\vspace{-0.1cm}
\end{framed}
\end{comment}

% !TeX root = ../main.tex

\subsection{Privacy Compliance Re-inspection}
%\section{A Longitudinal Analysis}
\label{sec:longitudinal}

%\paragraph{Re-visit the Tested SDK Libraries}
%Our study aims to unveil the comprehensive landscape of privacy compliance at the SDK level. 
%Instead of conducting a one-shot analysis, we carry out a re-inspection to explore the trend of privacy collection among the same group of the tested SDKs.
On the basis of our large-scale analysis performed on \allsdk SDKs collected in October 2022, we re-visited the two sources and collected the same group of SDKs in their latest release version as of October 2023 and conducted a cross-version analysis.
Table~\ref{tab:table-taint-analysis-stat-longitudinal} depicts a summary of our re-inspection results.
 
% !TeX root = ../../main.tex

\newcommand{\blueunchanged}{\scalerel*{\includegraphics{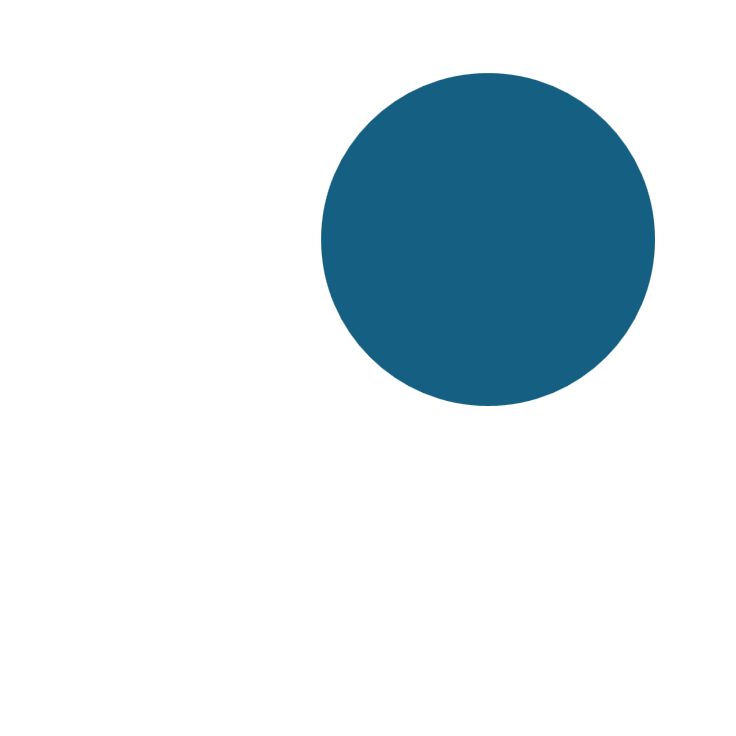}}{up}\xspace} % Show a blue color dot
\newcommand{\redup}{\scalerel*{\includegraphics{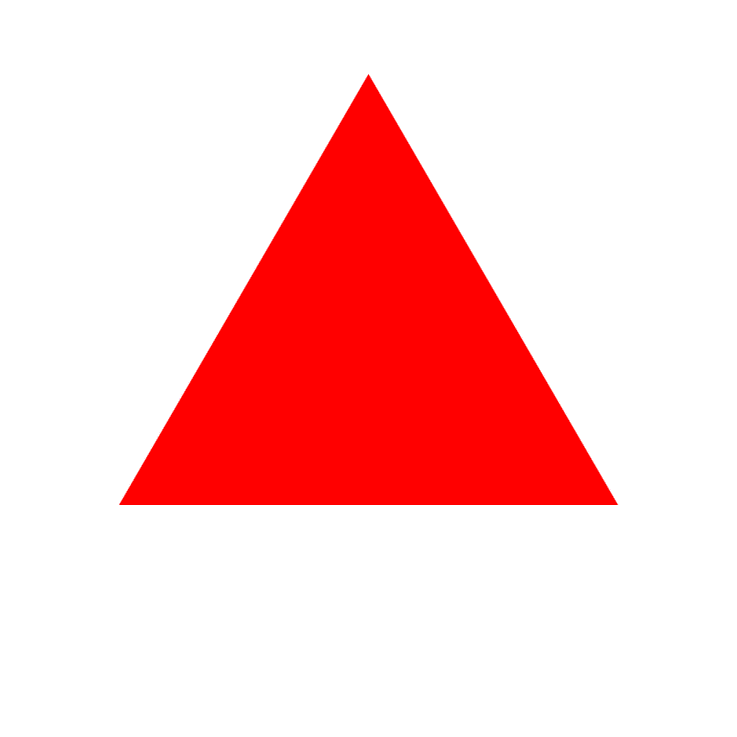}}{up}\xspace} % Show a red color up arrow
\newcommand{\greendown}{\scalerel*{\includegraphics{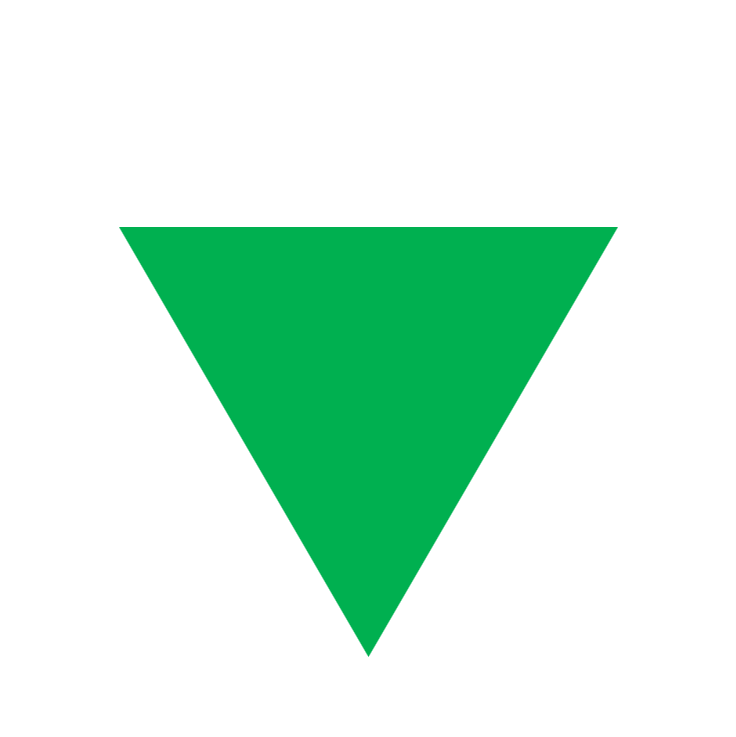}}{down}\xspace} % Show a green color down arrow

\begin{table}[t]
\centering
% Config the row spacing
\def\arraystretch{1.02}
% Config the column spacing
\setlength{\tabcolsep}{1pt}
\caption{\label{tab:table-taint-analysis-stat-longitudinal} Changes in the tainted traces after 12 months (only tainted data types are displayed)}
\vspace{-0.1cm}
\footnotesize

\begin{tabular}{llr|H|llr}
%\begin{tabular}{>{\arraybackslash}p{0.6\hsize}>{\raggedright\arraybackslash}X{0.3\hsize}}
\hline
\multicolumn{2}{l}{\textbf{\makecell[lt]{Data type \&\\class}}} & \textbf{\makecell[rt]{Changes in\\tainted traces}} & & \multicolumn{2}{l}{\textbf{\makecell[lt]{Data type \&\\class}}} & \textbf{\makecell[rt]{Changes in\\tainted traces}} \\ \hline
\textbf{C1} & Carrier info & \redup +31 (new)\textsuperscript{$\dagger$}	& & \textbf{C2} & \textit{Subtotal} & \greendown \textit{-106 (-61\%)} \\\cline{5-7}
& IMEI & \greendown -17 (65\%)			& & \textbf{C3} & Camera & \redup +2 (new)\\
& MEID & \greendown -2 (50\%)			& &  & Location & \greendown -94 (72\%)\\
& IMSI & \greendown -2 (40\%)			& &  & Misc sensors & \redup +20 (new)\\
& ICCID & \blueunchanged +0 (0\%)		& &  & \textit{Subtotal} & \greendown \textit{-72 (55\%)} \\ \cline{5-7}
& Serial & \redup +3 (60\%)				& & \textbf{C4} & Android ID & \redup +34 (new)\\
& SIM card info & \redup +1 (new)		& &  & App list & \redup +53 (new)\\
& SMS & \redup +4 (new)					& &  & Audio record & \redup +5 (new)\\
& Telephone no & \redup +2 (new)		& &  & Google Ad ID & \redup +9 (new)\\
& \textit{Subtotal} & \redup \textit{+20 (45\%)}	& &  & OAID & \redup +3 (new)\\ \cline{1-3}
\textbf{C2} & Bluetooth & \redup +5 (new)		& &  & Screen record & \redup +1 (new)\\
& Bluetooth MAC & \redup +3 (new)		& &  & \textit{Subtotal} & \redup \textit{+105 (new)}\\ \cline{5-7}
& BSSID/SSID & \greendown -73 (68\%)	& & \textbf{C5} & Clipboard data & \redup +17 (new)\\
& IP Address & \redup +9 (900\%)		& &  & Contact list & \redup +1 (new)\\
& WiFi  & \redup +9 (new)				& &  & \textit{Subtotal} & \redup \textit{+18 (new)} \\ \cline{5-7}
& WiFi MAC & \greendown -59 (92\%)		& & \multicolumn{2}{l}{\textbf{Total}}  &  \greendown -35 (10\%)\\ \hline
\end{tabular}
\vspace{-0.2cm}
\begin{flushleft}
	\begin{footnotesize}
		\textsuperscript{$\dagger$} We use ``new'' to indicate data types observed re-inspection for the first time.\\
	\end{footnotesize}
\end{flushleft}\vspace{-0.3cm} 
\end{table}

%\begin{table}[t]
%\caption{\label{tab:table-taint-analysis-stat-longitudinal} Statistic of the tainted traces from our longitudinal analysis (only tainted data types are displayed)}
%\vspace{-0.3cm}
% trim order: <left> <lower> <right> <upper>}
%\includegraphics[trim=0cm 0cm 0cm 0cm,clip=true,width=1\linewidth]{images/taint_statistic_longitudinal_single_col.pdf}
%\vspace{-0.5cm}
%\end{table}

\paragraph{Trends of Privacy Collection Behaviors}
%We assess the same group of SDKs in their latest release versions and summarize the tainted traces in Table~\ref{tab:table-taint-analysis-stat-longitudinal}. 
We find \taintallsdknew traces tainted from the latest version of \allsdk SDKs. 
Compared with our initial assessment, we observed 35 (10.1\%) fewer traces that attempt to collect and share user privacy data in public spaces.

The change between the outcomes of the two assessments mainly falls in the sources of tainted traces.
As shown in Table~\ref{tab:table-taint-analysis-stat-longitudinal}, we find obvious growth from the sharing of data relevant to media and installed software (C4), reserving over 1/3 of the total caught traces.
Among the 105 tainted traces, reading and sharing the list of installed apps ranks the top among all sources, with 53 traces caught. 
Although the access to data in that category has been pervasively observed in our initial assessment, our re-inspection did not catch any traces of \emph{sharing} such data. 

Additionally, we find that the collection behaviors for location and SSID/BSSID have significantly declined, marking the two greatest decreases. There is also an overall reduction in the collection of UUIs, including data types in categories C1 and C2.
Upon reviewing the historical changes in permission requirements, we note that most data types showing decreased collection behaviors had their access rules significantly updated in Android 10~\cite{google2023privacy10}, which imposed strict restrictions on third-party apps' access to personally identifiable information.
In contrast, we find the growth of tainted traces for the app list. Although the access to app list has also been restricted in a future release, these results still seriously concern us since the latest version of SDKs implies their persistent attempt to excessively and stealthily collect user privacy data.

\begin{comment}
We also observe that the trends in SDKs' privacy compliance show \emph{hysteresis} along with the evolution of Android OS, given Android 10 was released in 2019.
Such hysteresis could also be reflected in the data type demonstrating the largest growth of tainted traces, namely the app list.
We consider SDK developers to treat the app list as an alternative to the UUIs and share it for user identification and profiling purposes.
However, we find the SDK developers' efforts in collecting user privacy lag behind Google's strengthening privacy protection mechanism of Android OS, since access to the app list has been restricted since Android 11, which is just one year after the release of Android 10~\cite{google2023privacy11}.
In other words, SDK's privacy collection via the app list is infeasible on most devices nowadays. 
Nonetheless, given the number of tainted traces does not experience a significant change between the two assessments, SDK developers' privacy compliance still seriously concerns us since their latest implementation still implies their persistent attempt to excessively and stealthily collect users' privacy data.
\end{comment}

%\input{sections/responsible-disclosure}
% !TeX root = ../main.tex

\section{System Settings Injection Issues}
\label{sec:issues}

\begin{figure*}[t!]
\centering
% trim order: <left> <lower> <right> <upper>}
\includegraphics[trim=0cm 13.65cm 0cm 0cm,clip=true,width=0.75\linewidth]{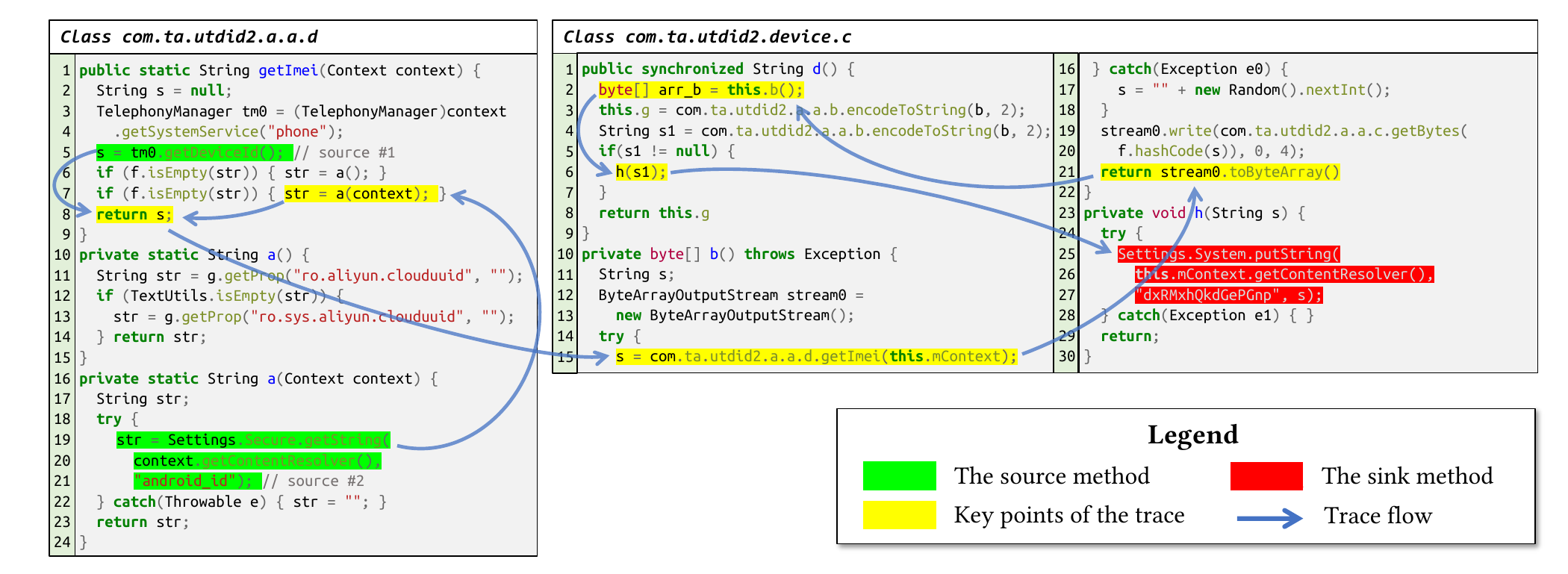}
%\vspace{-0.25cm}
\caption{Information flow demonstrates how the device ID is accessed and saved to the system settings by one of our analyzed SDKs (the obfuscated code is obtained from reverse engineering and has been simplified to save space)}
\label{fig:ali-sdk}
%\vspace{-0.6cm}
\end{figure*}

% !TeX root = ../../main.tex
\begin{table}[t]
\centering
% Config the row spacing
\def\arraystretch{1.1}
% Config the column spacing
\setlength{\tabcolsep}{2pt}
\caption{\label{tab:table-exploits} System setting entries found that contains a UUI}
\vspace{-0.1cm}
\footnotesize
\begin{tabular}{>{\arraybackslash}p{0.13\hsize}
		>{\arraybackslash}p{0.33\hsize}
		>{\arraybackslash}p{0.48\hsize}}
	\hline
	\textbf{\makecell[lt]{SDK}} & \textbf{\makecell[lt]{Entry name}} & \textbf{\makecell[lt]{Description of the value}}  \\\hline
	\multirow[t]{2}{*}{\makecell[lt]{Alicloud\\Push}} & \texttt{dxCRMxhQkdGePGnp}      & 
	IMEI/MEID, serial, or Android ID\textsuperscript{$\dagger$}, AES encrypted, base64 encoded    \\ \cline{2-3}
	& \texttt{mqBRboGZkQPcAkyk} & IMEI/MEID, serial, or Android ID\textsuperscript{$\dagger$}, AES encrypted, base64 encoded\\ \hline
	\multirow[t]{2}{*}{\makecell[lt]{Baidu\\LBS}}     & \texttt{com.baidu.deviceid.v2} & Package name and Android ID, MD5 hashed\textsuperscript{$\ddagger$}  \\  \cline{2-3}
	& \texttt{com.baidu.uuid}        & Package name and serial or a random UUID\textsuperscript{$\dagger$}, MD5 hashed\textsuperscript{$\ddagger$}   \\\hline
	
\end{tabular}
\begin{flushleft}
	\begin{footnotesize}
		\textsuperscript{$\dagger$} depending on the permissions granted to the embedded app. \\
            \textsuperscript{$\ddagger$} inferred description based on the reading logic in the implementation. \\
	\end{footnotesize}
\end{flushleft}\vspace{-0.2cm}
\end{table}
%\vspace{-0.1cm}

During our analysis, we find some traces from an SDK named ``Alicloud push'' are sunk at writing a value into the system settings.
Specifically, two entries of system settings are involved, as shown in Table~\ref{tab:table-exploits}.
Writing into the system settings requires privileged permission so that ordinary third-party apps are unable to do so.
However, we find that SDKs being embedded by pre-installed apps
%\footnote{We provide more technical details of the pre-installed apps and permission for writing into system settings in Appendix~\ref{sec:appendix-privilege}.} 
can write arbitrary values into the system settings even without the user's awareness.
We name this finding a \emph{system settings injection issue}, because it can be manipulated by an SDK to create and share user-unresettable identifiers (UUIs).

Although leaking privacy data through covert channels has been studied in existing literature~\cite{reardon2019fifty}, the system settings injection issues still seriously concern us as the caught SDK subtly evades accessing highly sensitive UUIs (e.g., IMEI, as discussed in ~\cite{reardon2019fifty}) but attempts to generate UUIs on their own. 
The caught SDK then takes advantage of its popularity and abuses the ``\texttt{preinstalled}'' privilege by counting on itself to be embedded in pre-installed apps or even privileged system apps.
By exploiting this issue, the system settings become a \emph{covert channel} to enable privacy-sensitive data sharing to apps without privileged permission.

Figure~\ref{fig:ali-sdk} illustrates an example of system setting injection from our findings. 
The source code is obtained through reverse engineering. 
In order to obtain a UUI, it attempts to invoke a privileged API (line 5, left), query the existence of pre-written system properties (lines 6 and 10-15, left), and read the Android ID (lines 7 and 16-24, left) sequentially, as shown in the figure. 
Due to the first approach requests the embedded app to be a system app with ``\texttt{READ\_PRIVILEGED\_PHONE\_STATE}'' permission, and the second approach is only viable when there exists a system app on the same device to write the two system properties in advance, we consider both them \emph{infeasible} by default on nowadays Android devices. 
However, the third approach, i.e., reading Android ID through a system setting table named \texttt{SECURE}, does not request special permission. 
As a result, the SDK can at least obtain the Android ID to uniquely identify the device. It then performs a hash function over the obtained Android ID (lines 19-20, right) and writes the hash value into the system settings with a customized key (lines 25-27, right).  
This process, although it complies with the Android permission mechanism, transforms the value of \emph{resettable} Android ID into a \emph{de facto} UUI that can be accessed by any app that knows the customized key in the system settings.

Considering the exploitability of the pre-installed privileges, we carry out another round of investigation of all \allsdk SDKs to find if there exists any other potential privacy infringement by taking advantage of the system settings mechanism.
As a result, we managed to find another two entries that have been read in an SDK named ``Baidu LBS'' (see Table~\ref{tab:table-exploits}). 
As their names imply, these two entries are obviously written by third-party privileged apps. 
We suspect the formation of the two entries should follow a similar routine, and accordingly reverse engineer the involved SDK and present our inferred descriptions of the two entries in Table~\ref{tab:table-exploits}.
Unfortunately, we did not find any app to write these entries into the system settings.
Who exploited this system settings issue remains unknown to us.

\paragraph{Impact} 
We notice that the two studied SDKs are widely adopted in third-party apps, especially in the Chinese market. 
Therefore, we conduct a small-scale study to explore the impact of the system settings injection issues. 
We collect devices from 9 OEM manufacturers (8 OEM devices purchased in the Chinese market\footnote{These devices are manufactured by Honor, Huawei, Lenovo, OnePlus, Oppo, Samsung, Vivo, and Xiaomi.} and 1 Google Pixel device purchased in the US market), download the packages of pre-install apps, and investigate whether the two aforementioned SDKs have been once embedded in the pre-installed apps.
We observe a worrisome fact that except for the Google Pixel devices, all the tested devices have at least one pre-installed app using the investigated SDKs.
According to the latest market report~\cite{statcounter2024worldwide}, these 8 OEM manufacturers reserve over 66\% global Android smartphone market share. They also shipped over 94\% new Android devices in China in 2023~\cite{statcounter2024china}.
This implies that almost all Android smartphones sold in the Chinese market may already pose a privacy leakage risk based on an assumption that the choice of pre-installed apps should remain the same across different  models produced by a manufacturer.

\section{Discussion}
\label{sec:discussion}
\subsection{Causes of Privacy Mishandling}
\paragraph{Privilege Free-riding}
An SDK's access to privacy data depends on the permissions granted to the apps it's embedded in. However, third-party apps pre-installed by OEMs can obtain privileged permissions. 
A malicious SDK in such an app can exploit system settings to deploy privilege free-riding attacks, creating and sharing UUIs without users' awareness. As reading system settings require no permissions, privilege free-riding significantly undermines Google's efforts in restricting third-party app access to UUIs since Android 10.

\paragraph{Insufficient Regulations}
It is well known that the release of Android apps is subject to strict regulations, with non-compliance in privacy preservation potentially leading to removal from app stores or fines. However, SDK releases lack similar fine-grained regulations and sufficient legislation to ensure compliance with personal data protection laws.
Ironically, even some Google in-house SDKs do not have privacy policies listed on the SDK Index\footnote{An example can be found at \url{https://play.google.com/sdks/details/com-google-android-gms-play-services-auth}}. Currently, both repositories function merely as search engines for available SDKs, without a platform-level compliance checking mechanism to ensure complete permission disclosure.

\subsection{Our Recommendations}
%\textcolor{blue}{We recommend addressing the privilege abuse attack surface created by pre-installed apps from the following three aspects, where malicious SDKs can create UUIs and share them across installed apps. First, Google should restrict pre-installed apps from writing to system settings. OEMs should minimize pre-installed apps. }

\paragraph{Precautions of Privilege Abuse}
We regard that the pre-installed privileges open an attack surface for a malicious SDK, specifically when the SDK uses app-differential Android ID to identify a device user, transforms it into a de facto UUI, and provides free access with all apps installed on the victim device. 
Based on this finding, we advocate Google restricting pre-installed apps from writing data into the system settings in the future.
% In fact, pre-installed apps are free to create a customized UUI and share it even if it has not read any privacy data.
%Stop automatically granting \texttt{WRITE\_SETTINGS} and instead allow apps with ``\texttt{preinstalled}'' privilege to explicitly request it at runtime can mitigate the risk of system settings injection.
We also advise OEM manufacturers to minimize the number of pre-installed apps on their devices. 

%\textcolor{blue}{Second, strengthened platform regulation is needed, as current SDK release platforms lack substantial enforcement. We advocate comprehensive regulations like Google's App Store, including static analysis of SDK binaries to exhaustively list requested permissions, addressing the black-box concern for app developers. }

\paragraph{Strengthened Platform Regulation}
Currently, both release platforms selected in this study are merely index websites, which do not host the SDK binaries and lack substantial enforcement.
Developers voluntarily register SDKs on these platforms.  
For that reason, we call for strengthened regulation over the usage of Android SDKs. 
%Most potential privacy leakage issues can be identified through code analysis, and therefore the risk of privacy infringement can be significantly mitigated if the release platforms such as Google Play SDK Index enforce strict and comprehensive regulations, like what has been deployed in its App Store. 
%SDKs containing malicious code should be flagged to warn app developers.
We also advocate platforms performing static analysis over the SDK binaries to exhaustively list requested permissions, especially for those at the dangerous level and above. This measure could address the black-box concern for app developers.

%\textcolor{blue}{Third, transparent permission requests from embedded SDKs would allow developers discretion in granting permissions without compromising functionality. Google's new \emph{SDK Runtime} in Android 13 aims to improve SDK governance, but adoption remains challenging given many SDKs are released outside Google platforms~\cite{google2024sdk}.}

\paragraph{Transparent App Development}
We advocate transparent management over the permission requested by the embedding SDKs. Thus, app developers can have their own discretion in granting permissions without compromising functionality.
We are delighted to see Google's efforts in improving the SDK's governance. A new mechanism called \emph{SDK Runtime} has been introduced in Android 13 that provides a dedicated execution environment for SDKs~\cite{google2024sdk}. However, this mechanism is not compulsory and still faces many usability and functionality limitations at the moment of this work being performed. Besides that, how to ensure the cooperation of SDK developers remains a challenging issue, given many SDKs are released and maintained in non-Google platforms.

\subsection{Limitations}
\paragraph{Absence of Dynamic Testing}
While our static taint analysis approach is effective, dynamic testing can improve efficiency by automatically identifying and excluding logically invalid tainted traces, reducing manual validation efforts. However, dynamic SDK testing is challenging due to inadequate documentation, undisclosed interfaces, and highly event-driven/credential-required logic. For example, some finance and customer service SDKs require a valid license or token to function. Satisfying all functional prerequisites of the collected SDKs can be another challenging task and therefore is out of the scope of this study. We plan to explore potential dynamic testing solutions in the future.

\paragraph{Coverage of Targeted Methods}
Our analysis covered 741 API source methods and 46 sinks.
Although we frequently update the configuration during this study, we may still miss some data exfiltration channels as new APIs can be introduced along with OS evolution and emerging hardware. 
%We frequently update the configuration, but newly introduced methods accessing or sharing privacy data could be missed initially.

\paragraph{Elusive and Ambiguous Disclosure}
Although developers should clearly disclose personal data collection, we find that elusive or ambiguous phrasing widely exists in many SDK's privacy policies, leading to potential false negatives in our NLI model-based analysis. 
We aim to explore the latest advancements in LLMs and improve our approach to minimize such impacts.

%\paragraph{Scope of SDKs}
%Our assessment covered 158 Android SDKs from two mainstream release platforms. However, privacy data leakage can occur through non-Android APIs and even pure Java logic, suggesting our future work should extend beyond Android SDKs to any libraries or tools integrated into Android apps capable of handling personal data.

%Despite these limitations, our large-scale analysis provides crucial insights into the prevalence and risks of privacy data mishandling by SDKs, highlighting the need for stronger platform regulations, transparent permission requests, and improved SDK governance mechanisms like Android 13's SDK Runtime.
% !TeX root = ../main.tex

%\section{Responsible Disclosure}
\section{Ethics Considerations}
\label{sec:appendix-responsible-disclosure}
%\paragraph{Responsible Disclosure}
We have reported all our findings to the relevant parties and we also have kept them confidential for at least 90 days.
For each SDK that found a compliance issue, we manually searched the contact left by its developers and e-mailed our findings since our first round of SDK investigation.
Although we observed slight mitigation from re-inspection, we did not receive positive responses regarding non-compliance concerns from the developers that we have attempted to contact. 
We also expressed our concern to Google about the system settings injection issues (to be detailed in Section~\ref{sec:issues}) and suggested removing the pre-installed privileges in writing system settings in the future release.
%However, Google replied that the pre-installed privilege is not regarded as a vulnerability and the relevant permission mechanism would remain unchanged.
%It is at the device manufacturer's discretion to select and grant pre-installed privileges to third-party apps.
Google replied and suggested that it is at the device manufacturer's discretion to select and grant pre-installed privileges to third-party apps.

%\section{Open Science}
%We will open source all artifacts involved in this study upon paper acceptance, including a dataset of Android SDKs, our code for privacy policy inference, the configuration scripts for the static analysis, and the detailed statistics of our assessment results. 
\section{Related Works}
%We review prior works in the area of Android privacy compliance evaluation and third-party SDK privacy leakage analysis.

%With the blowout growth of mobile applications, privacy issues have emerged as a substantial concern for mobile device users. Therefore, privacy analysis regarding mobile applications have been exhaustively examined in recent years.
\paragraph{Privacy Compliance Evaluation}
Numerous works focus on detecting inconsistencies between the privacy practices of an application and its descriptions or its privacy policies~\cite{bui2023detection,gorla2014chabada,liu2018large,pandita2013whyper,qu2014autocog,slavin2016toward,wang2018guileak,zhou2022policycomp}. 
This line of research begins with automatic risk assessment of Android apps, such as WHYPER~\cite{pandita2013whyper} and AutoCog~\cite{qu2014autocog} that leverage NLP techniques to evaluate the necessity of required permissions based on apps' descriptions. Gorla et al.~\cite{gorla2014chabada} study mismatches between app descriptions and app behaviors by clustering Android apps based on their description topics and identifying outliers in each cluster with respect to their API usage. 
Recent work extends the scope of privacy compliance evaluation to examining third-party data harvesting, user consent, service misconfigurations, and quality of privacy policies at the app level. XFinder~\cite{wang2021understanding} uses dynamic analysis and NLP techniques to parse terms-of-service and extract data-sharing policies. Nguyen et al.~\cite{nguyen2021share} study user consent verification through traffic analysis and ablation experiments. Zhang et al.~\cite{zhang2020does} highlight the issue of analytical service misconfigurations, while Yu et al.~\cite{yu2018ppchecker} scrutinize the apps' privacy policies with regard to the inclusion of third-party libraries. Andow et al.~\cite{andow2020actions} propose POLICHECK that identifies ambiguous or omitted disclosures of third parties. 
Pan et al.~\cite{pan2024trap} study the automated generation of privacy policies for mobile apps and unveil that defective compliance widely exists in the automatically generated privacy policies.
Zhao et al.~\cite{zhao2023demystifying} study privacy compliance of third-party libraries in Android through the lens of pattern matching in privacy disclosure. Compared with it, our work investigates SDKs, which are more complex in implementation and business logic. Our study also resorts to a language model in privacy disclosure analysis that does not request manual data annotation and model training.
Parallel research by Xiao et al.~\cite{xiao2024measuring} investigate compliance of third-party libraries by assessing the consistency between API documents and dynamic analysis results.  
%Despite these efforts, existing tools often neglect to assess the compliance of third-party libraries themselves with regulatory requirements.
In comparison, our work focuses on compliance of SDKs that contains more complicated implementation that hinders from effective dynamic analysis, and meanwhile our study examine the privacy disclosure by directly interpreting the privacy policies, which are written in natural language without a standard format. 
%, checking the practical end-to-end data handling against declarations of developers. 
%We also specifically consider the elevated risk of privacy leakage in the context of pre-installed apps. 

\paragraph{Privacy Leakage Assessment}
There is a substantial body of literature addressing various aspects of privacy concerns in the Android ecosystem~\cite{enck2014taintdroid,zhang2009peep,zhou2013identity}. Meng et al.~\cite{meng2023post} study user-unresettable identifier safeguards on a wide range of Android devices. 
He et al.~\cite{he2018dynamic} utilize dynamic analysis to explore the leakage of permission-related data from third-party libraries in Android apps
Ekambaranathan et al.~\cite{ekambaranathan2021money} investigate data usage and disclosure in children's app. Liu et al.~\cite{liu2019privacy} examine data leakage from nine analytic libraries across 300 apps, using both static and dynamic analyses. Razaghpanah et al.~\cite{razaghpanah2018apps} detect third-party advertising and tracking services via dynamic analysis of network traffic data. 
%However, it identifies limited types of personal data and hardware identifiers.
%Our work comprehensively investigates privacy compliance at the SDK level on a large scale, complementing existing research to unveil the privacy preservation landscape in the software supply chain of the Android ecosystem.

% !TeX root = ../main.tex
\section{Conclusion}
In this study, we investigate the intricate privacy challenges associated with the widespread use of third-party SDKs in Android apps. Our large-scale analysis involved an evaluation of \allsdk third-party SDKs, prevalent across Android applications, with an approach intertwining rigorous taint analysis for data collection practices and LLM-aided interpretation of privacy policies. 
The findings were startling, with \correcttaintallsdk potential privacy leakages detected from \allsdk SDKs.
Besides that, our study reveals that less than 70\% of examined SDKs provide privacy policies, among which approximately 37\% SDKs are found indulging in the over-collection of privacy data, signaling a clear violation of privacy norms.
Our findings suggest the need for stricter regulations, improved development guidelines, and more transparent permission management to reduce privacy risks in Android apps.

\bibliographystyle{IEEEtran}
\bibliography{IEEEabrv,reference}

\begin{thebibliography}{10}
\providecommand{\url}[1]{#1}
\csname url@samestyle\endcsname
\providecommand{\newblock}{\relax}
\providecommand{\bibinfo}[2]{#2}
\providecommand{\BIBentrySTDinterwordspacing}{\spaceskip=0pt\relax}
\providecommand{\BIBentryALTinterwordstretchfactor}{4}
\providecommand{\BIBentryALTinterwordspacing}{\spaceskip=\fontdimen2\font plus
\BIBentryALTinterwordstretchfactor\fontdimen3\font minus
  \fontdimen4\font\relax}
\providecommand{\BIBforeignlanguage}[2]{{%
\expandafter\ifx\csname l@#1\endcsname\relax
\typeout{** WARNING: IEEEtran.bst: No hyphenation pattern has been}%
\typeout{** loaded for the language `#1'. Using the pattern for}%
\typeout{** the default language instead.}%
\else
\language=\csname l@#1\endcsname
\fi
#2}}
\providecommand{\BIBdecl}{\relax}
\BIBdecl

\bibitem{authe_lib}
\BIBentryALTinterwordspacing
P.~Ruiz, ``{Authenticating on Android with the AppAuth Library},'' 2022,
  (accessed 9 July 2024). [Online]. Available:
  \url{https://medium.com/androiddevelopers/authenticating-on-android-with-the-appauth-library-7bea226555d5}
\BIBentrySTDinterwordspacing

\bibitem{encryption_lib}
\BIBentryALTinterwordspacing
{Android Documentation}, ``{Work with data more securely},'' 2023, (accessed 9
  July 2024). [Online]. Available:
  \url{https://developer.android.com/topic/security/data}
\BIBentrySTDinterwordspacing

\bibitem{analytics_lib}
\BIBentryALTinterwordspacing
{Segment}, ``{Analytics for Android},'' 2023, (accessed 9 July 2024). [Online].
  Available:
  \url{https://segment.com/docs/connections/sources/catalog/libraries/mobile/android}
\BIBentrySTDinterwordspacing

\bibitem{advertisement_lib}
\BIBentryALTinterwordspacing
{Android Documentation}, ``{Add Ads to Your Instant App},'' 2023, (accessed 9
  July 2024). [Online]. Available:
  \url{https://developer.android.com/topic/google-play-instant/guides/advertising}
\BIBentrySTDinterwordspacing

\bibitem{UI_lib}
\BIBentryALTinterwordspacing
------, ``{Develop UI for Android},'' 2023, (accessed 9 July 2024). [Online].
  Available: \url{https://developer.android.com/develop/ui}
\BIBentrySTDinterwordspacing

\bibitem{salza2018}
P.~Salza, F.~Palomba, D.~Di~Nucci, C.~D'Uva, A.~De~Lucia, and F.~Ferrucci,
  ``{Do Developers Update Third-Party Libraries in Mobile Apps?}'' in
  \emph{ICPC}, 2018, p. 255–265.

\bibitem{zhang2018detecting}
Y.~Zhang, J.~Dai, X.~Zhang, S.~Huang, Z.~Yang, M.~Yang, and H.~Chen,
  ``Detecting third-party libraries in android applications with high precision
  and recall,'' in \emph{2018 IEEE 25th International Conference on Software
  Analysis, Evolution and Reengineering (SANER)}, 2018, pp. 141--152.

\bibitem{priv_leak_blackkite}
\BIBentryALTinterwordspacing
{blackkite}, ``{Data Breaches Caused By Third-Parties},'' 2022, (accessed 9
  July 2024). [Online]. Available:
  \url{https://blackkite.com/data-breaches-caused-by-third-parties}
\BIBentrySTDinterwordspacing

\bibitem{priv_leak_mahajan}
\BIBentryALTinterwordspacing
P.~Mahajan, ``{3rd Party Libraries: Your Next Data Breach Nightmare},'' 2022,
  (accessed 9 July 2024). [Online]. Available:
  \url{https://www.privado.ai/post/3rd-party-libraries-data-breach}
\BIBentrySTDinterwordspacing

\bibitem{htfma2023msm}
\BIBentryALTinterwordspacing
{Droid}, ``{How to fix my Android - What Is Mobile Services Manager?}'' 2023,
  (accessed 9 July 2024). [Online]. Available:
  \url{https://www.howtofixmyandroid.com/what-is-mobile-services-manager/}
\BIBentrySTDinterwordspacing

\bibitem{mollah2022mobile}
\BIBentryALTinterwordspacing
M.~Mollah, ``{What Is Mobile Services Manager? Is It A Threat? How To Fix
  It?}'' 2022, (accessed 9 July 2024). [Online]. Available:
  \url{https://www.socialmediamagazine.org/mobile-services-manager/}
\BIBentrySTDinterwordspacing

\bibitem{keegan2022sold}
\BIBentryALTinterwordspacing
J.~Keegan and A.~Ng, ``{Over 100 apps that sold location data to sketchy data
  broker revealed},'' 2022, (accessed 9 July 2024). [Online]. Available:
  \url{https://mashable.com/article/app-location-data-sold}
\BIBentrySTDinterwordspacing

\bibitem{gdpr2016}
{The European Parliament}, ``Regulation (eu) 2016/679 of the european
  parliament and of the council of 27 april 2016 on the protection of natural
  persons with regard to the processing of personal data and on the free
  movement of such data, and repealing directive 95/46/ec (general data
  protection regulation) (text with eea relevance),'' \emph{Official Journal of
  the European Union}, 2016.

\bibitem{ccpa2023california}
\BIBentryALTinterwordspacing
{State of California Department of Justice}, ``California consumer privacy act
  (ccpa),'' 2023, (accessed 9 July 2024). [Online]. Available:
  \url{https://oag.ca.gov/privacy/ccpa}
\BIBentrySTDinterwordspacing

\bibitem{google_play_sdk_index}
\BIBentryALTinterwordspacing
{Google}, ``Google play sdk index,'' 2024, (accessed 9 July 2024). [Online].
  Available: \url{https://play.google.com/sdks}
\BIBentrySTDinterwordspacing

\bibitem{caict}
\BIBentryALTinterwordspacing
{China Academy of Information and Communication Technology}, ``Nationwide sdk
  management and service platform (translated from chinese),'' 2024, (accessed
  9 July 2024). [Online]. Available: \url{https://sdk.caict.ac.cn/official}
\BIBentrySTDinterwordspacing

\bibitem{rodriguez2024sharing}
D.~Rodriguez, J.~M. Del~Alamo, C.~Fern{\'a}ndez-Aller, and N.~Sadeh, ``Sharing
  is not always caring: Delving into personal data transfer compliance in
  android apps,'' \emph{IEEE Access}, 2024.

\bibitem{google2024understanding}
\BIBentryALTinterwordspacing
Google, ``Understand app privacy \& security practices with google play's data
  safety section,'' 2024, (accessed 9 July 2024). [Online]. Available:
  \url{https://support.google.com/googleplay/answer/11416267?sjid=6208336964583751082-AP}
\BIBentrySTDinterwordspacing

\bibitem{wolford2024guide}
\BIBentryALTinterwordspacing
B.~Wolford, ``A guide to gdpr data privacy requirements,'' 2024, (accessed 9
  July 2024). [Online]. Available: \url{https://gdpr.eu/data-privacy/}
\BIBentrySTDinterwordspacing

\bibitem{google2024manifest}
\BIBentryALTinterwordspacing
Google, ``{Manifest.permission},'' 2024, (accessed 28 June 2024). [Online].
  Available:
  \url{https://developer.android.com/reference/android/Manifest.permission}
\BIBentrySTDinterwordspacing

\bibitem{google2023privacy10}
\BIBentryALTinterwordspacing
------, ``{Privacy changes in Android 10},'' 2023, (accessed 9 July 2024).
  [Online]. Available:
  \url{https://developer.android.com/about/versions/10/privacy/changes}
\BIBentrySTDinterwordspacing

\bibitem{chen2014information}
T.~Chen, I.~Ullah, M.~A. Kaafar, and R.~Boreli, ``Information leakage through
  mobile analytics services,'' in \emph{{HotMobile}}, 2014, pp. 1--6.

\bibitem{leung2016should}
C.~Leung, J.~Ren, D.~Choffnes, and C.~Wilson, ``Should you use the app for
  that? comparing the privacy implications of app-and web-based online
  services,'' in \emph{{IMC}}, 2016, pp. 365--372.

\bibitem{papadopoulos2017long}
E.~P. Papadopoulos, M.~Diamantaris, P.~Papadopoulos, T.~Petsas, S.~Ioannidis,
  and E.~P. Markatos, ``The long-standing privacy debate: Mobile websites vs
  mobile apps,'' in \emph{{WWW}}, 2017, pp. 153--162.

\bibitem{ren2018longitudinal}
J.~Ren, M.~Lindorfer, D.~J. Dubois, A.~Rao, D.~Choffnes, and
  N.~Vallina-Rodriguez, ``A longitudinal study of pii leaks across android app
  versions,'' in \emph{{NDSS}}, 2018.

\bibitem{ren2016recon}
J.~Ren, A.~Rao, M.~Lindorfer, A.~Legout, and D.~Choffnes, ``Recon: Revealing
  and controlling pii leaks in mobile network traffic,'' in \emph{{MobiSys}},
  2016, pp. 361--374.

\bibitem{wang2021understanding}
J.~Wang, Y.~Xiao, X.~Wang, Y.~Nan, L.~Xing, X.~Liao, J.~Dong, N.~Serrano,
  H.~Lu, X.~Wang \emph{et~al.}, ``Understanding malicious cross-library data
  harvesting on android,'' in \emph{{USENIX Security}}, 2021.

\bibitem{enck2014taintdroid}
W.~Enck, P.~Gilbert, S.~Han, V.~Tendulkar, B.-G. Chun, L.~P. Cox, J.~Jung,
  P.~McDaniel, and A.~N. Sheth, ``Taintdroid: an information-flow tracking
  system for realtime privacy monitoring on smartphones,'' \emph{{TOCS}},
  vol.~32, no.~2, pp. 1--29, 2014.

\bibitem{gibler2012androidleaks}
C.~Gibler, J.~Crussell, J.~Erickson, and H.~Chen, ``Androidleaks: automatically
  detecting potential privacy leaks in android applications on a large scale,''
  in \emph{{TrustCom}}.\hskip 1em plus 0.5em minus 0.4em\relax Springer, 2012,
  pp. 291--307.

\bibitem{qiu2015apptrace}
L.~Qiu, Z.~Zhang, Z.~Shen, and G.~Sun, ``Apptrace: Dynamic trace on android
  devices,'' in \emph{{ICC}}, 2015, pp. 7145--7150.

\bibitem{sun2016taintart}
M.~Sun, T.~Wei, and J.~C. Lui, ``Taintart: A practical multi-level
  information-flow tracking system for android runtime,'' in \emph{{CCS}},
  2016, pp. 331--342.

\bibitem{andow2020actions}
B.~Andow, S.~Y. Mahmud, J.~Whitaker, W.~Enck, B.~Reaves, K.~Singh, and
  S.~Egelman, ``Actions speak louder than words: {Entity-Sensitive} privacy
  policy and data flow analysis with {PoliCheck},'' in \emph{29th USENIX
  Security Symposium (USENIX Security 20)}, 2020, pp. 985--1002.

\bibitem{bui2021consistency}
D.~Bui, Y.~Yao, K.~G. Shin, J.-M. Choi, and J.~Shin, ``Consistency analysis of
  data-usage purposes in mobile apps,'' in \emph{Proceedings of the 2021 ACM
  SIGSAC Conference on Computer and Communications Security}, 2021, pp.
  2824--2843.

\bibitem{selenium}
\BIBentryALTinterwordspacing
B.~Muthukadan, ``{Selenium with Python},'' 2018, (accessed 9 July 2024).
  [Online]. Available: \url{https://selenium-python.readthedocs.io/}
\BIBentrySTDinterwordspacing

\bibitem{andow2019policylint}
B.~Andow, S.~Y. Mahmud, W.~Wang, J.~Whitaker, W.~Enck, B.~Reaves, K.~Singh, and
  T.~Xie, ``{PolicyLint}: investigating internal privacy policy contradictions
  on google play,'' in \emph{28th USENIX security symposium (USENIX security
  19)}, 2019, pp. 585--602.

\bibitem{xie2022scrutinizing}
F.~Xie, Y.~Zhang, C.~Yan, S.~Li, L.~Bu, K.~Chen, Z.~Huang, and G.~Bai,
  ``Scrutinizing privacy policy compliance of virtual personal assistant
  apps,'' in \emph{{ASE}}, 2022.

\bibitem{yu2016can}
L.~Yu, X.~Luo, X.~Liu, and T.~Zhang, ``Can we trust the privacy policies of
  android apps?'' in \emph{2016 46th Annual IEEE/IFIP International Conference
  on Dependable Systems and Networks (DSN)}.\hskip 1em plus 0.5em minus
  0.4em\relax IEEE, 2016, pp. 538--549.

\bibitem{zimmeck2016automated}
S.~Zimmeck, Z.~Wang, L.~Zou, R.~Iyengar, B.~Liu, F.~Schaub, S.~Wilson,
  N.~Sadeh, S.~Bellovin, and J.~Reidenberg, ``Automated analysis of privacy
  requirements for mobile apps,'' in \emph{2016 AAAI Fall Symposium Series},
  2016.

\bibitem{harkous2022hark}
H.~Harkous, S.~T. Peddinti, R.~Khandelwal, A.~Srivastava, and N.~Taft, ``Hark:
  A deep learning system for navigating privacy feedback at scale,'' in
  \emph{2022 IEEE Symposium on Security and Privacy (SP)}.\hskip 1em plus 0.5em
  minus 0.4em\relax IEEE, 2022, pp. 2469--2486.

\bibitem{rasthofer2014machine}
S.~Rasthofer, S.~Arzt, and E.~Bodden, ``A machine-learning approach for
  classifying and categorizing android sources and sinks.'' in \emph{NDSS},
  vol.~14, 2014, p. 1125.

\bibitem{google2023loginfo}
\BIBentryALTinterwordspacing
Google, ``{Log Info Disclosure},'' 2023, (accessed 9 July 2024). [Online].
  Available:
  \url{https://developer.android.com/topic/security/risks/log-info-disclosure}
\BIBentrySTDinterwordspacing

\bibitem{google2022sandbox}
\BIBentryALTinterwordspacing
------, ``{Application Sandbox},'' 2022, (accessed 9 July 2024). [Online].
  Available: \url{https://source.android.com/docs/security/app-sandbox}
\BIBentrySTDinterwordspacing

\bibitem{meng2023post}
M.~H. Meng, Q.~Zhang, G.~Xia, Y.~Zheng, Y.~Zhang, G.~Bai, Z.~Liu, S.~G. Teo,
  and J.~S. Dong, ``Post-gdpr threat hunting on android phones: dissecting
  os-level safeguards of user-unresettable identifiers,'' in \emph{{NDSS}},
  2023.

\bibitem{arzt2014flowdroid}
S.~Arzt, S.~Rasthofer, C.~Fritz, E.~Bodden, A.~Bartel, J.~Klein, Y.~Le~Traon,
  D.~Octeau, and P.~McDaniel, ``Flowdroid: Precise context, flow, field,
  object-sensitive and lifecycle-aware taint analysis for android apps,''
  \emph{Acm Sigplan Notices}, vol.~49, no.~6, pp. 259--269, 2014.

\bibitem{mariana_trench}
\BIBentryALTinterwordspacing
{Facebook Open Source}, ``{Mariana Trench: Security-Focused Static Analysis for
  Android and Java Applications},'' 2024, (accessed 9 July 2024). [Online].
  Available: \url{https://mariana-tren.ch/}
\BIBentrySTDinterwordspacing

\bibitem{appshark}
\BIBentryALTinterwordspacing
{Bytedance}, ``{Appshark},'' 2022, (accessed 9 July 2024). [Online]. Available:
  \url{https://github.com/bytedance/appshark}
\BIBentrySTDinterwordspacing

\bibitem{shen2021understanding}
Y.~Shen, P.-A. Vervier, and G.~Stringhini, ``Understanding worldwide private
  information collection on android,'' in \emph{{NDSS}}, 2021.

\bibitem{wang2020exploring}
Z.~Wang, Z.~Li, M.~Xue, and G.~Tyson, ``Exploring the eastern frontier: A first
  look at mobile app tracking in china,'' in \emph{Passive and Active
  Measurement: 21st International Conference, PAM 2020, Eugene, Oregon, USA,
  March 30--31, 2020, Proceedings 21}.\hskip 1em plus 0.5em minus 0.4em\relax
  Springer, 2020, pp. 314--328.

\bibitem{reardon2019fifty}
J.~Reardon, {\'A}.~Feal, P.~Wijesekera, A.~E.~B. On, N.~Vallina-Rodriguez, and
  S.~Egelman, ``50 ways to leak your data: An exploration of apps'
  circumvention of the android permissions system,'' in \emph{{USENIX
  Security}}, 2019.

\bibitem{statcounter2024worldwide}
\BIBentryALTinterwordspacing
{Statcounter}, ``{Mobile Vendor Market Share Worldwide},'' 2024, (accessed 9
  July 2024). [Online]. Available:
  \url{https://gs.statcounter.com/vendor-market-share/mobile}
\BIBentrySTDinterwordspacing

\bibitem{statcounter2024china}
\BIBentryALTinterwordspacing
------, ``{Mobile Vendor Market Share China},'' 2024, (accessed 9 July 2024).
  [Online]. Available:
  \url{https://gs.statcounter.com/vendor-market-share/mobile/china}
\BIBentrySTDinterwordspacing

\bibitem{google2024sdk}
\BIBentryALTinterwordspacing
Google, ``{SDK Runtime},'' 2024, (accessed 9 July 2024). [Online]. Available:
  \url{https://developer.android.com/design-for-safety/privacy-sandbox/sdk-runtime}
\BIBentrySTDinterwordspacing

\bibitem{bui2023detection}
D.~Bui, B.~Tang, and K.~G. Shin, ``Detection of inconsistencies in privacy
  practices of browser extensions,'' in \emph{2023 IEEE Symposium on Security
  and Privacy (SP)}.\hskip 1em plus 0.5em minus 0.4em\relax IEEE, 2023, pp.
  2780--2798.

\bibitem{gorla2014chabada}
A.~Gorla, I.~Tavecchia, F.~Gross, and A.~Zeller, ``Checking app behavior
  against app descriptions,'' in \emph{{ICSE}}, 2014.

\bibitem{liu2018large}
X.~Liu, Y.~Leng, W.~Yang, W.~Wang, C.~Zhai, and T.~Xie, ``A large-scale
  empirical study on android runtime-permission rationale messages,'' in
  \emph{{VL/HCC}}, 2018.

\bibitem{pandita2013whyper}
R.~Pandita, X.~Xiao, W.~Yang, W.~Enck, and T.~Xie, ``Whyper: Towards automating
  risk assessment of mobile applications,'' in \emph{{USENIX Security}}, 2013,
  pp. 527--542.

\bibitem{qu2014autocog}
Z.~Qu, V.~Rastogi, X.~Zhang, Y.~C. Chen, T.~Z. Zhu, and Z.~Chen, ``Autocog:
  Measuring the description-to-permission fidelity in android applications,''
  in \emph{{CCS}}, 2014.

\bibitem{slavin2016toward}
R.~Slavin, X.~Wang, M.~B. Hosseini, J.~Hester, R.~Krishnan, J.~Bhatia, T.~D.
  Breaux, and J.~Niu, ``Toward a framework for detecting privacy policy
  violations in android application code,'' in \emph{Proceedings of the 38th
  International Conference on Software Engineering}, 2016, pp. 25--36.

\bibitem{wang2018guileak}
X.~Wang, X.~Qin, M.~B. Hosseini, R.~Slavin, T.~D. Breaux, and J.~Niu,
  ``Guileak: Tracing privacy policy claims on user input data for android
  applications,'' in \emph{Proceedings of the 40th International Conference on
  Software Engineering}, 2018, pp. 37--47.

\bibitem{zhou2022policycomp}
L.~Zhou, C.~Wei, T.~Zhu, G.~Chen, X.~Zhang, S.~Du, H.~Cap, and H.~Zhu,
  ``{POLICYCOMP}: Counterpart comparison of privacy policies uncovers overbroad
  personal data collection practices,'' in \emph{{USENIX Security}}, 2022.

\bibitem{nguyen2021share}
T.~T. Nguyen, M.~Backes, N.~Marnau, and B.~Stock, ``Share first, ask later (or
  never?) studying violations of $\{$GDPR's$\}$ explicit consent in android
  apps,'' in \emph{30th USENIX Security Symposium (USENIX Security 21)}, 2021,
  pp. 3667--3684.

\bibitem{zhang2020does}
X.~Zhang, X.~Wang, R.~Slavin, T.~Breaux, and J.~Niu, ``How does
  misconfiguration of analytic services compromise mobile privacy?'' in
  \emph{Proceedings of the ACM/IEEE 42nd International Conference on Software
  Engineering}, 2020, pp. 1572--1583.

\bibitem{yu2018ppchecker}
L.~Yu, X.~Luo, J.~Chen, H.~Zhou, T.~Zhang, H.~Chang, and H.~K. Leung,
  ``Ppchecker: Towards accessing the trustworthiness of android apps’ privacy
  policies,'' \emph{IEEE Transactions on Software Engineering}, vol.~47, no.~2,
  pp. 221--242, 2018.

\bibitem{pan2024trap}
S.~Pan, D.~Zhang, M.~Staples, Z.~Xing, J.~Chen, X.~Xu, and T.~Hoang, ``Is it a
  trap? a large-scale empirical study and comprehensive assessment of online
  automated privacy policy generators for mobile apps,'' in \emph{33rd USENIX
  Security Symposium (USENIX Security 24)}, 2024, pp. 5681--5698.

\bibitem{zhao2023demystifying}
K.~Zhao, X.~Zhan, L.~Yu, S.~Zhou, H.~Zhou, X.~Luo, H.~Wang, and Y.~Liu,
  ``Demystifying privacy policy of third-party libraries in mobile apps,'' in
  \emph{2023 IEEE/ACM 45th International Conference on Software Engineering
  (ICSE)}.\hskip 1em plus 0.5em minus 0.4em\relax IEEE, 2023, pp. 1583--1595.

\bibitem{xiao2024measuring}
Y.~Xiao, C.~Zhang, Y.~Qin, F.~F.~S. Alharbi, L.~Xing, and X.~Liao, ``Measuring
  compliance implications of third-party libraries' privacy label disclosure
  guidelines,'' in \emph{Proceedings of the 2024 on ACM SIGSAC Conference on
  Computer and Communications Security}, 2024, pp. 1641--1655.

\bibitem{zhang2009peep}
K.~Zhang and X.~Wang, ``Peeping tom in the neighborhood: Keystroke
  eavesdropping on multi-user systems,'' in \emph{{USENIX Security}}, 2009.

\bibitem{zhou2013identity}
X.~Zhou, S.~Demetriou, D.~He, M.~Naveed, X.~Pan, X.~Wang, C.~A. Gunter, and
  K.~Nahrstedt, ``Identity, location, disease and more: Inferring your secrets
  from android public resources,'' in \emph{Proceedings of the 2013 ACM SIGSAC
  conference on Computer \& communications security}, 2013, pp. 1017--1028.

\bibitem{he2018dynamic}
Y.~He, B.~Hu, and Z.~Han, ``Dynamic privacy leakage analysis of android
  third-party libraries,'' in \emph{2018 1st International Conference on Data
  Intelligence and Security (ICDIS)}, 2018, pp. 275--280.

\bibitem{ekambaranathan2021money}
A.~Ekambaranathan, J.~Zhao, and M.~Van~Kleek, ````money makes the world go
  around'': Identifying barriers to better privacy in children's apps from
  developers' perspectives,'' in \emph{Proceedings of the 2021 CHI Conference
  on Human Factors in Computing Systems}, 2021, pp. 1--15.

\bibitem{liu2019privacy}
X.~Liu, J.~Liu, S.~Zhu, W.~Wang, and X.~Zhang, ``Privacy risk analysis and
  mitigation of analytics libraries in the android ecosystem,'' \emph{IEEE
  Transactions on Mobile Computing}, vol.~19, no.~5, pp. 1184--1199, 2019.

\bibitem{razaghpanah2018apps}
A.~Razaghpanah, R.~Nithyanand, N.~Vallina-Rodriguez, S.~Sundaresan, M.~Allman,
  C.~Kreibich, P.~Gill \emph{et~al.}, ``Apps, trackers, privacy, and
  regulators: A global study of the mobile tracking ecosystem,'' in \emph{The
  25th Annual Network and Distributed System Security Symposium (NDSS 2018)},
  2018.

\end{thebibliography}

%\appendix
%\input{sections/appendix}

\end{document}